\documentclass[twocolumn]{aastex631}

\usepackage{xspace}
\usepackage{xcolor}
\usepackage{amsmath}
\usepackage{longtable}
\usepackage{showyourwork}

\DeclareUnicodeCharacter{2212}{\textendash}

\newcommand{\taucz}{$\tau_\mathrm{cz}$\xspace}
\newcommand{\rocrit}{$\mathrm{Ro_{crit}}$\xspace}
\newcommand{\rothresh}{$\mathrm{Ro_{thresh}}$\xspace}
\newcommand{\rosun}{$\mathrm{Ro_{\odot}}$\xspace}
\newcommand{\rocritfinal}{$\mathrm{Ro_{crit}}~\lesssim~\mathrm{Ro_\odot}$\xspace}
\newcommand{\lamostkep}{LAMOST--Kepler\xspace}
\newcommand{\lamostmcq}{LAMOST--McQuillan\xspace}
\newcommand{\lamostsan}{LAMOST--Santos\xspace}
\newcommand{\teffmin}{5500~K\xspace}
\newcommand{\agerange}{$\sim$2--6~Gyr\xspace}

\newcommand{\jvs}{vS19\xspace}
\newcommand{\mma}{MMA14\xspace}
\newcommand{\hall}{H21\xspace}
\newcommand{\masuda}{MPH21\xspace}
\newcommand{\curtis}{C20\xspace}
\newcommand{\santos}{S21\xspace}

\newcommand{\teff}{\ensuremath{T_{\mathrm{eff}}}\xspace}  
\newcommand{\logg}{\ensuremath{\log g}\xspace} 

\newcommand{\vsini}{\ensuremath{v \sin i}\xspace} 
\newcommand{\msun}{$M_\odot$\xspace}

\newcommand{\tsun}{$T_\mathrm{eff,\odot}$\xspace}

\newcommand{\rearth}{$R_\oplus$\xspace}
\newcommand{\prot}{\ensuremath{P_\mathrm{rot}}\xspace}

\newcommand{\rper}{\ensuremath{R_\mathrm{per}}\xspace}

\newcommand{\cca}{Center for Computational Astrophysics, Flatiron Institute, New York, NY 10010, USA}
\newcommand{\amnh}{Department of Astrophysics, American Museum of Natural History, Central Park West at 79th Street, New York, NY 10024, USA}
\newcommand{\columbia}{Department of Astronomy, Columbia University, 550 West 120th Street, New York, NY, USA}
\newcommand{\ifa}{Institute for Astronomy, University of Hawai’i, Honolulu, HI, USA}

\received{February 4, 2022}
\revised{April 20, 2022}
\accepted{May 5, 2022}

\submitjournal{ApJ}

\shorttitle{The Rossby Ridge}
\shortauthors{David et al.}

\graphicspath{{./}{figures/}}

\begin{document}
\title{Further Evidence of Modified Spin-down in Sun-like Stars:\\Pileups in the Temperature--Period Distribution}

\correspondingauthor{Trevor J. David}
\email{tdavid@flatironinstitute.org}

\author[0000-0001-6534-6246]{Trevor J.\ David}
\affil{\cca}
\affil{\amnh}

\author[0000-0003-4540-5661]{Ruth Angus}
\affil{\amnh}
\affil{\cca}
\affil{\columbia}

\author[0000-0002-2792-134X]{Jason L.~Curtis}
\affil{\columbia}

\author[0000-0002-4284-8638]{Jennifer L.~van Saders}
\affil{\ifa}

\author[0000-0001-8196-516X]{Isabel L.~Colman}
\affil{\amnh}

\author[0000-0002-3011-4784]{Gabriella Contardo}
\affil{\cca}

\author[0000-0003-4769-3273]{Yuxi Lu}
\affil{\columbia}
\affil{\amnh}

\author[0000-0002-7550-7151]{Joel C.~Zinn}
\affil{\amnh}

\begin{abstract}
We combine stellar surface rotation periods determined from NASA's Kepler mission with spectroscopic temperatures to demonstrate the existence of pileups at the long-period and short-period edges of the temperature--period distribution for main-sequence stars with temperatures exceeding $\sim$\teffmin. The long-period pileup is well-described by a curve of constant Rossby number, with a critical value of \rocritfinal. The long-period pileup was predicted by \citet{vanSaders2019} as a consequence of weakened magnetic braking, in which wind-driven angular momentum losses cease once stars reach a critical Rossby number. Stars in the long-period pileup are found to have a wide range of ages (\agerange), meaning that, along the pileup, rotation period is strongly predictive of a star's surface temperature but weakly predictive of its age. The short-period pileup, which is also well-described by a curve of constant Rossby number, is not a prediction of the weakened magnetic braking hypothesis but may instead be related to a phase of slowed surface spin-down due to core–envelope coupling. The same mechanism was proposed by \citet{Curtis2020} to explain the overlapping rotation sequences of low-mass members of differently aged open clusters. The relative dearth of stars with intermediate rotation periods between the short- and long-period pileups is also well-described by a curve of constant Rossby number, which aligns with the period gap initially discovered by \citet{McQuillan2013b} in M-type stars. These observations provide further support for the hypothesis that the period gap is due to stellar astrophysics, rather than a non-uniform star-formation history in the Kepler field.
\end{abstract}


\section{Introduction} \label{sec:intro}
Solar-type and low-mass stars ($M\lesssim1.3$~\msun) lose mass and angular momentum through magnetized winds \citep{Parker1958, WeberDavis1967, Mestel1968, Kawaler1988}. Consequently, stellar rotation rates are observed to decline with age. \citet{Skumanich1972} presented the first attempt to calibrate this age-rotation relationship using the rotation periods of Sun-like stars in open clusters with independently determined ages, finding a $P_\mathrm{rot} \propto t^{1/2}$ scaling, where $t$ is stellar age. In the intervening decades, observational determinations of stellar rotation periods among open cluster members revealed how stellar spin rates evolve in more detail, leading to the calibration of the so-called gyrochronology method \citep{Barnes2003, Barnes2007, Barnes2010, MamajekHillenbrand2008, Meibom2009, Angus2019}.

The arrival of continuous, high-precision, long-baseline photometry from NASA's Kepler space telescope \citep{Borucki2010} provided a watershed moment for stellar rotation studies, yielding period detections for tens of thousands of stars \citep[e.g.][]{Reinhold2013, McQuillan2014, Santos2021} and allowing for gyrochronology to be extended to older ages \citep[e.g.][]{Meibom2011, Meibom2015}. NASA's subsequent K2 \citep{Howell2014} and TESS \citep{Ricker2015} missions propelled the field of stellar rotation further still, providing an exquisitely detailed picture of how spin rates evolve for stars with a broad range of masses and ages in stellar associations \citep[e.g.][]{Douglas2016, Douglas2017, Douglas2019, Rebull2016, Rebull2017, Rebull2018, Rebull2020, Curtis2019a, Curtis2019b, Curtis2020}. New and evermore precise data is becoming available at a rate that is outpacing efforts to re-calibrate gyrochronology, which is necessary to capture the complex relationship between a star's spin and its age. 

For example, efforts to calibrate gyrochronology relations using Kepler asteroseismic targets revealed tension with relations calibrated to open clusters and found that rotation periods could not be described by a single power-law relation with age \citep{Angus2015}. This tension, at least in part, is due to the fact that standard gyrochronology models are unable to account for the anomalously rapid rotation rates of stars older than the Sun, leading to the suggestion that stars with Rossby numbers of Ro~$\gtrsim$~\rosun experience a phase of weakened magnetic braking \citep[WMB,][]{vanSaders2016}. 

Forward modeling simulations of the observed Kepler rotation period distribution also provided support for the WMB hypothesis over standard spin-down models, in that WMB models are better able to match the observed long-period edge \citep[][hereafter vS19]{vanSaders2019}. Those authors also predicted a pileup of stars along the long-period edge, which they hypothesized could not be seen in the \citet{McQuillan2014} sample (hereafter \mma) due to large errors on \teff in the revised Kepler Input Catalog \citep[KIC,][]{Huber2014}. While \jvs favored the WMB hypothesis to explain observations, those authors were also careful to point out that a long-period edge can be caused by detection biases, as stars with longer rotation periods (and larger Rossby numbers) have smaller amplitude variations which pose more difficulty to period-detection algorithms. 

More recently, \citet{Hall2021}, hereafter H21, used the asteroseismic rotation rates of Kepler dwarfs, with different selection and detection biases\footnote{Only 48/91 stars in the \hall sample had rotation periods that were also detected from rotational brightness modulations.} from the \jvs study and the present work, to argue support for the WMB model. \citet{Masuda2021}, hereafter MPH21, also found support for the WMB hypothesis from inference of the rotation period distribution of Sun-like stars using stellar radii and projected rotational velocities.  While the physics responsible for the weakened magnetic braking of solar-type stars is unknown, one hypothesis is that the declining efficiency of wind-driven angular momentum loss is connected to the magnetic field complexity, which may vary with Rossby number \citep[e.g.][]{Reville2015, vanSaders2016, Garraffo2016, Metcalfe2016, Metcalfe2019}.  

Here we examine the rotation period distribution of stars observed by Kepler, leveraging the recent release of precise spectroscopic parameters from large-scale surveys, to demonstrate the existence of pileups at the long- and short-period edges of the \teff--\prot distribution of solar-type stars. We discuss our sample in \S\ref{sec:sample}, describe the steps of our analysis in \S\ref{sec:analysis}, discuss some implications of these results in \S\ref{sec:discussion}, and present our conclusions in \S\ref{sec:conclusions}.

\section{Sample Selection} \label{sec:sample}
\begin{figure*}
    \centering
    \includegraphics[width=\textwidth]{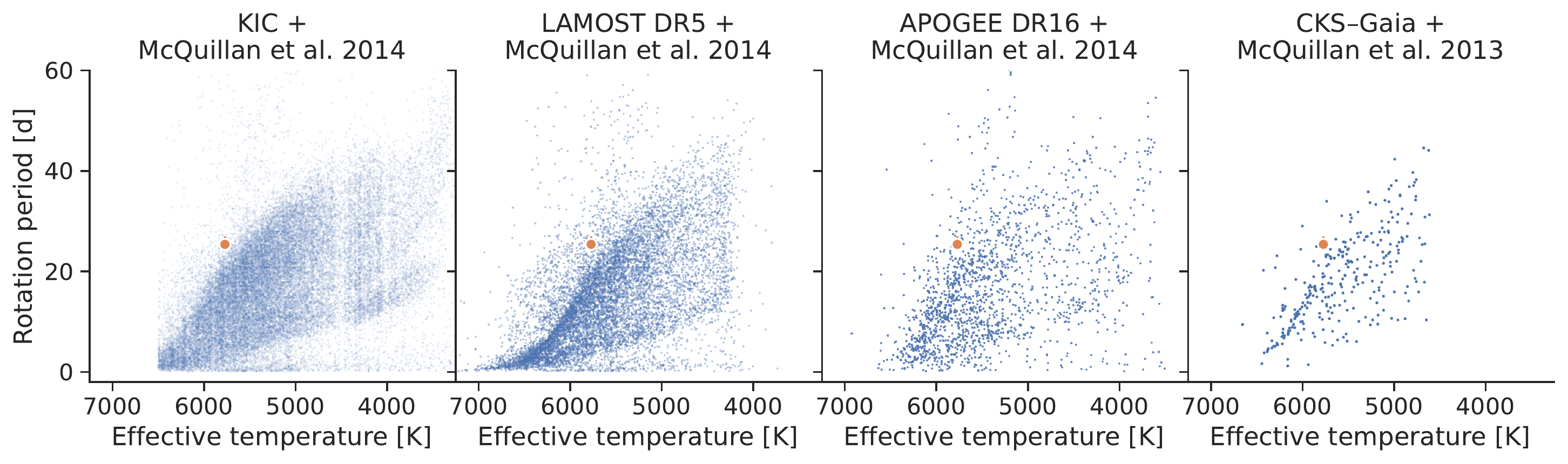}
    \caption{The \teff-\prot plane using rotation periods from \mma or, in the case of the CKS sample, \citet{McQuillan2013} which applied an identical analysis to Kepler Objects of Interest (KOIs), with \teff originating from the source denoted at top. The \mma \teff values originate from the Kepler Input Catalog \citep[KIC,][]{Brown2011} or \citet{Dressing2013} for low-mass stars. The orange point in each panel indicates the Sun's equatorial rotation period, with the errorbar capturing the range of periods measured from the activity belts. Many of the stars above the long-period pileup are subgiants which have experienced spin-down due to expansion off the main-sequence, as pointed out in \jvs.}
    \label{fig:surveys}
    \script{surveys.py}
\end{figure*}

Below, we describe the samples utilized in this work. All stars characterized here were targets of NASA's Kepler mission \citep{Borucki2010} and have published rotation periods derived from the Kepler data. For each subsample of the Kepler field described below, we combined published rotation periods from a variety of literature sources with spectroscopic parameters provided by large-scale surveys (Figure~\ref{fig:surveys}).

\subsection{California--Kepler Survey} \label{subsec:cks}
The CKS project acquired high-resolution spectroscopy for 1305 Kepler planet host stars \citep{Petigura2017}. CKS spectra were acquired with the Keck/HIRES spectrograph \citep{Vogt1994} and spectroscopic parameters were determined by averaging parameters from the SpecMatch pipeline \citep{Petigura2015} and SME@XSEDE, a Python implementation of the Spectroscopy Made Easy pipeline \citep{Valenti1996}. The internal (relative) errors on \teff from the CKS catalog are estimated at $\pm$60~K, with systematic errors of $\pm$100~K estimated from comparison to other catalogs \citep[see Table 7 of][]{Petigura2017}. The metallicity distribution of the CKS sample is centered near solar, with a mean and standard deviation of +0.03~dex and 0.18~dex, respectively.

We compiled rotation periods for these stars from a variety of literature sources including \citet{McQuillan2013, Mazeh2015} and \citet{Angus2018}. For each star in the sample we then visually inspected the Kepler light curve folded on all available literature periods, as well as the first harmonics and sub-harmonics of those periods, and recorded our preferred period along with a reliability flag. Our procedure is explained in detail in \S2.1 of \citet{David2021}, and rotation period vetting sheets for each Kepler Object of Interest (KOI) are publicly available through Zenodo.\footnote{\url{http://10.0.20.161/zenodo.4645437}} The vast majority of stars in the CKS sample host small planets ($R_P < 4$~\rearth) and as such it is not expected that the host stars have experienced tidal spin-up from the planets.

In addition to the original CKS catalog, we also cross-matched our sample with the catalogs of \citet{Brewer2018} and \citet{Martinez2019}, both of which presented spectroscopic parameters for CKS stars based on independent analysis of the same spectra. The \citet{Brewer2018} study, referred to here as SPOCS, also published elemental abundances and ages from isochrone fitting for the CKS sample. We additionally cross-matched the CKS catalog with the LAMOST DR5 catalog \citep{Xiang2019} which is described further in \S\ref{subsec:lamost}. We compare the \teff--\prot distributions of the CKS sample using \teff and rotation periods from a variety of sources in Appendix~\ref{app:teffprot}.

\subsection{LAMOST \label{subsec:lamost}}
The LAMOST project derived homogeneous spectroscopic parameters from low-resolution ($R\sim$~1800) LAMOST DR5 spectra for approximately 40\% of the Kepler field \citep{Zong2018, Xiang2019}. A description of the LAMOST Kepler field observations is provided in \citet{deCat2015}. The \citet{Xiang2019} catalog derived stellar parameters using the DD-Payne pipeline, which builds on the method of \citet{Ting2017b} by incorporating elements of the Cannon \citep{Ness2015} and uses the overlap with GALAH DR2 and APOGEE DR14 as training data. We cross-matched the LAMOST DR5 stellar parameter catalog of \citet{Xiang2019} with the \mma catalog, which published rotation periods for $>$34000 Kepler targets, as well as the \citet{Santos2021} catalog (hereafter S21), which provides rotation periods for $>$55000 FGKM stars observed by Kepler.

We matched 10844 LAMOST targets to 10550 unique Kepler IDs in \mma, resulting in a sample with well-determined \prot and spectroscopic \teff (having a median error of 24~K). For the Kepler sources with duplicate cross-matched LAMOST sources we kept the source with a brighter Gaia DR2 $G$ magnitude. In the \santos sample we found 54982 unique cross-matched sources in LAMOST, of which 18990 have published temperatures and rotation periods. The metallicity distribution of the LAMOST--McQuillan sample is centered near solar, with a mean and standard deviation of $-0.1$~dex and 0.26~dex, respectively. There is negligible overlap (3 stars) between our LAMOST--McQuillan sample and the CKS sample since \mma did not publish rotation periods for KOIs, which were the targets of the CKS project. Rotation periods for KOIs are instead published in \citet{McQuillan2013}, as discussed in \S\ref{subsec:cks}. Visualizations of the \lamostmcq and \lamostsan samples are shown in Figure~\ref{fig:xmatch}.

\newpage
\subsection{APOGEE}
The Apache Point Observatory Galactic Evolution Experiment \citep[APOGEE;][]{Majewski2017} is a large-scale, high-resolution ($R \sim 22500$) stellar spectroscopic survey conducted at $H$-band as part of the Sloan Digital Sky Survey \citep[SDSS-IV;][]{Blanton2017}. The spectroscopic analysis pipeline for SDSS DR16 is described in \citet{Jonsson2020}. We used Gaia DR2 source IDs \citep{Gaia2016, Gaia2018} to cross-match Megan Bedell's Gaia--Kepler catalog\footnote{\url{https://gaia-kepler.fun/}} with the APOGEE DR16 catalog \citep{Ahumada2020}. Kepler IDs were then used to cross-match this table with the \mma catalog. While the current overlap between Kepler targets and APOGEE is small compared to the LAMOST catalog, APOGEE DR17 will contain more dwarf stars and provide a better resource for studies such as ours. The focus of this work are overdensities in the \teff--\prot plane, and as these appear to be less prominent when using  APOGEE DR16 temperatures (Figure~\ref{fig:surveys}) we conclude that LAMOST and CKS provide more precise estimates of \teff and do not analyze the APOGEE sample further. 

\subsection{The Sun}
To place the Sun in the context of the long-period pileup, we use up-to-date estimates of the Sun's effective temperature, color, rotation period, and age. Following the IAU 2015 Resolution B3 \citep{Prsa2016}, we take the nominal effective temperature of the Sun to be $\mathcal{T}^\mathrm{N}_\mathrm{eff,\odot} = 5772$~K. For the color and apparent magnitude of the Sun in the Gaia bandpasses we adopt $G_\mathrm{BP}-G_\mathrm{RP} = 0.818$~mag and $G = -26.895$~mag \citep{Casagrande2018}. For the solar rotation period we adopt the equatorial rotation period of $P_\mathrm{eq,\odot}~\approx~25.4$~d. While helioseismic differential rotation studies have inferred rotation periods of up to $\approx$~36~d near the poles \citep[][and references therein]{Thompson2003}, sunspots are rarely observed outside the activity belts, roughly $\pm 30^\circ$ from the equator \citep[][and references therein]{Hathaway2015}. We therefore adopt 27~d as an approximate upper limit to the rotation period of the Sun, as it would be observed by Kepler through rotational brightness modulations.\footnote{We note that the \jvs models used in \S\ref{subsec:models} are calibrated to the Sun's equatorial rotation period of 25.4~d.} The age of the Sun is assumed to be 4.567~Gyr from Pb-Pb dating of calcium-aluminum inclusions and chondrules recovered from primitive meteorites \citep[][and references therein]{Bahcall1995}.

\begin{figure*}
    \centering
    \includegraphics[width=\textwidth]{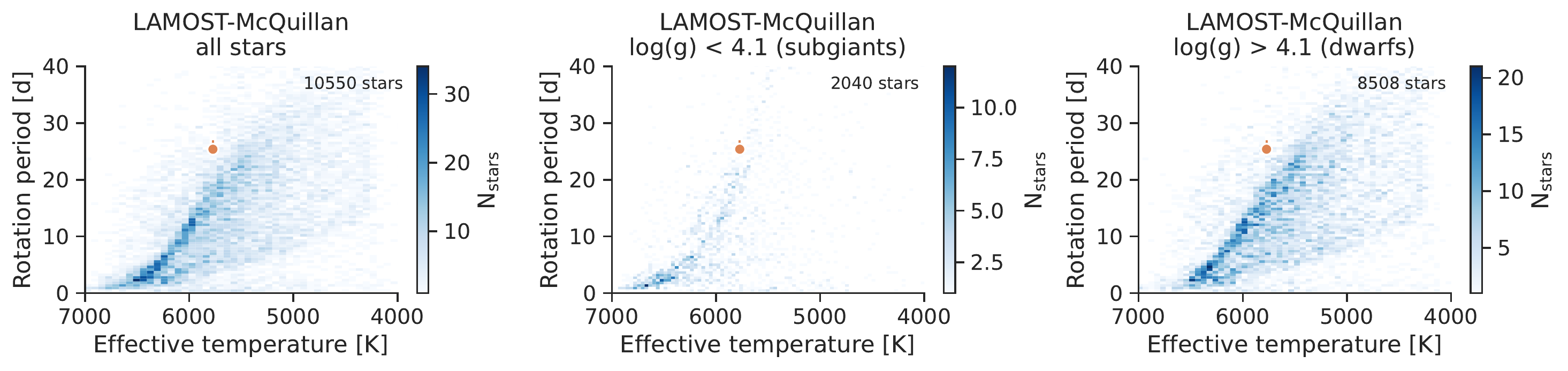}
    \includegraphics[width=\textwidth]{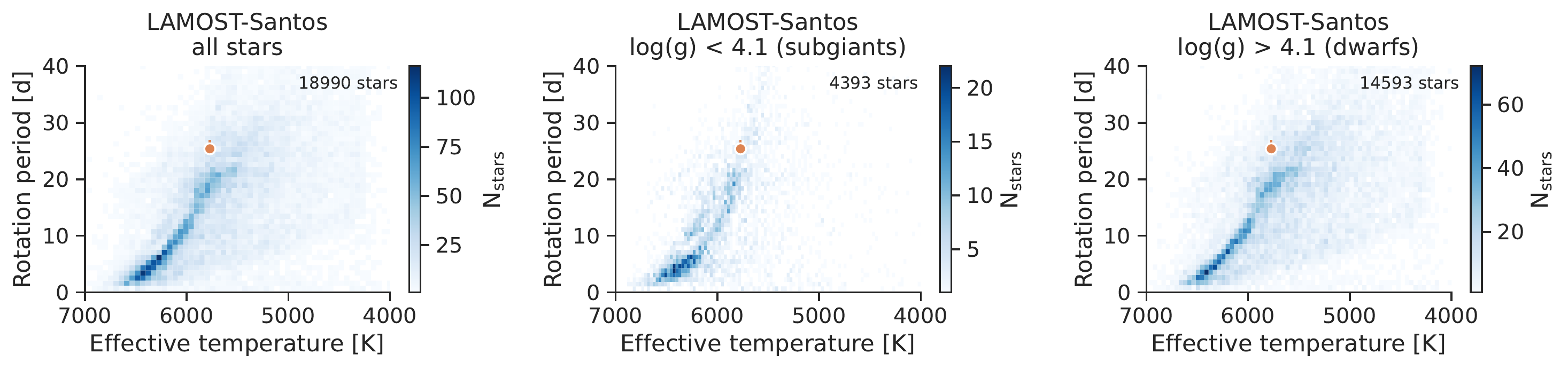}
    \caption{Two-dimensional histograms indicating the number of Kepler target stars in the \teff–\prot plane for the \lamostmcq sample (top row) and the \lamostsan sample (bottom row). The effects of a simple cut in \logg to separate subgiants and dwarfs are shown in the middle and right columns. In each panel the Sun is indicated by the orange point, with an errorbar reflecting the range of periods measured from the activity belts. The long-period pileup for dwarf stars is clearly seen to extend to the solar temperature. The short-period pileup is clearer in the smaller \lamostmcq dwarf sample, potentially because the \santos catalog detected more stars at longer periods. The secondary overdensity observed in the subgiant samples, most visible in the bottom center panel, appears to be at twice the period of the primary overdensity, potentially due to erroneously determined rotation periods.}
    \label{fig:xmatch}
    \script{xmatch.py}
\end{figure*}

\section{Analysis}
\label{sec:analysis}

\subsection{Initial observations}
We first noted a pileup of stars along the long-period edge for stars with \teff~$>5800$~K when examining the \teff-\prot plane for the CKS sample, using the CKS \teff values and the vetted rotation period compilation from \citet{David2021}. The pileup in that sample is visible even when sourcing \prot uniformly from \citet{McQuillan2013}, as shown in the rightmost panel of Figure~\ref{fig:surveys}. Sourcing rotation periods from \citet{Mazeh2015}, \citet{Angus2018}, and \santos revealed that this pileup is still apparent when adopting periods uniformly from other catalogs as well (see Appendix~\ref{app:teffprot}). It is thus the \teff precision afforded by spectroscopic catalogs that reveal the long-period pileup. In Appendix~\ref{app:gaia}, we demonstrate that it is possible to recover the long-period pileup in the color–period plane using high-precision photometry from the Gaia mission \citep{Gaia2016} and selecting stars with minimal interstellar reddening.

\begin{figure*}
    \centering
    \includegraphics[width=\linewidth]{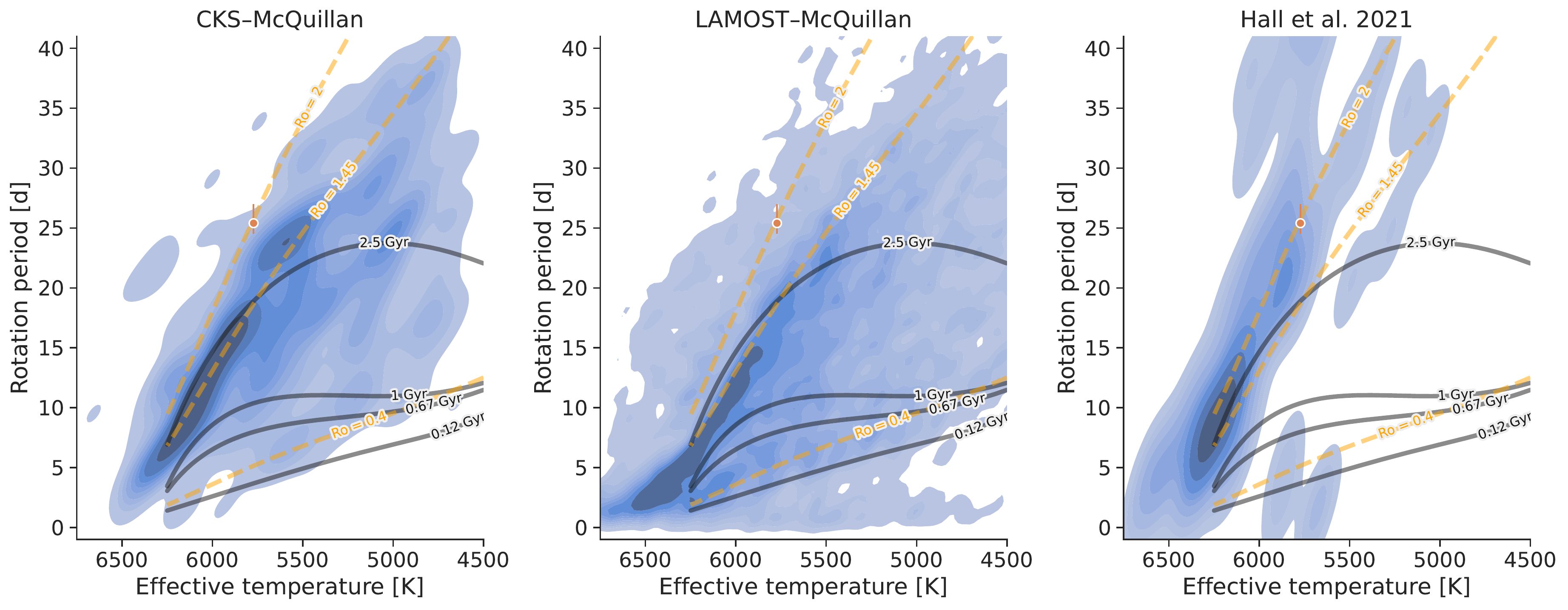}
    \caption{Gaussian kernel density estimation (blue contours) of the \teff--\prot distributions of the CKS--McQuillan, \lamostmcq, and asteroseismic \hall samples, from left to right. Empirical cluster sequences from \curtis are shown by the dark grey lines. The orange dashed lines show constant Rossby curves of fiducial values (see \S\ref{subsec:rossby}). The short-period pileup can be observed in the LAMOST--McQuillan sample for \teff~$\gtrsim$~5500~K. The orange point indicates the Sun's temperature and equatorial rotation period, with the errorbar capturing the range of periods measured from the activity belts.}
    \label{fig:kde}
    \script{kde.py}
\end{figure*}

Comparing the long-period pileup to rotation period sequences from open clusters can provide insight into the ages of stars on the pileup. The long-period pileup is close to the empirical hybrid cluster sequence derived by \citet{Curtis2020}, hereafter \curtis, from members of the NGC~6819 (age~$\sim$~2.5~Gyr) and Ruprecht~147 (age~$\sim$~2.7~Gyr) open clusters (Figure~\ref{fig:kde}). Notably, we use the color--\teff relation presented by those authors to recast the cluster sequences in terms of \teff.  The observation that the long-period pileup approximately corresponds with the $\sim$2.5--2.7~Gyr cluster sequence implies that stars with \teff$\gtrsim$~\teffmin have already piled up onto the edge by or before this timescale. Similarly, the long-period pileup clearly lies at longer periods than the empirical $\sim$1~Gyr cluster sequence based on rotation rates in the NGC~6811 cluster \citep{Curtis2019a, Curtis2020}, implying that it takes F-type stars $>1$~Gyr to reach the long-period pileup. These observations are in accordance with predictions from the WMB model which suggest the pileup forms on a timescale of 2--3~Gyr \citep{vanSaders2019}. We show in \S\ref{subsec:ages} that stars along the pileup have a range of ages, with a lower bound that is consistent with $\sim$2~Gyr.

\begin{figure}
    \centering
    \includegraphics[width=\linewidth]{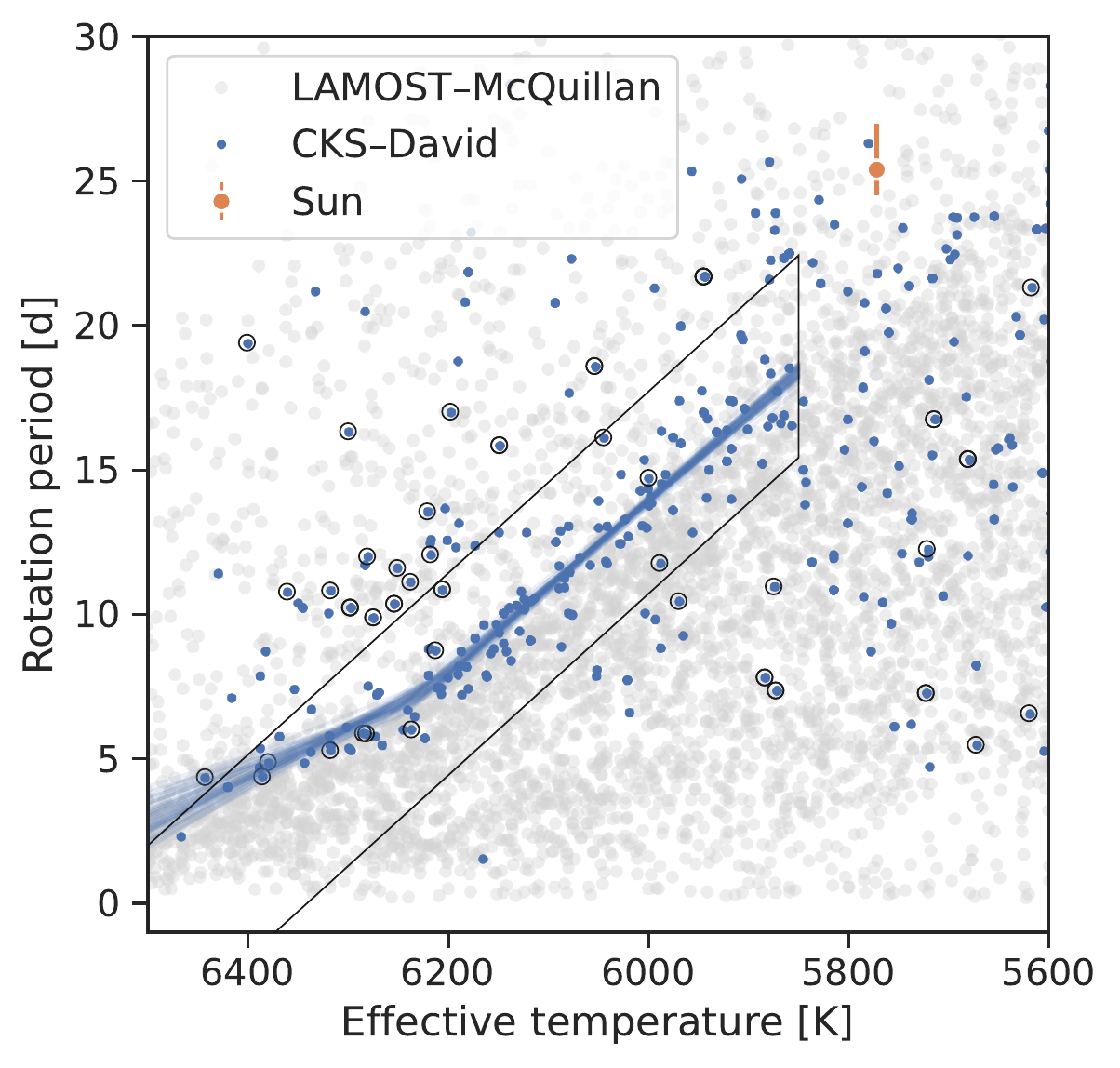}
    \caption{The \teff–\prot plane for the \lamostmcq sample (grey points) and the CKS sample (blue points) with periods sourced from \citet{David2021}. The selection used to identify long-period pileup stars in the CKS sample is shown by the region enclosed by the black lines. Stars with \teff differences $>100$~K between the CKS \citep{Fulton2018} and SPOCS \citep{Brewer2018} catalogs are shown by open black circles. The orange point indicates the Sun. Random draws from the MCMC chain of a piecewise linear fit to the CKS long-period pileup are shown by the blue lines. An inflection point in the long-period edge is apparent near \teff=~6250~K.}
    \label{fig:inflection}
    \script{inflection.py}
\end{figure}

A change in slope along the long-period edge is observed in both the CKS and LAMOST samples (Figure~\ref{fig:inflection}), with an inflection point corresponding closely to the Kraft break at $\approx$~6250~K, the point at which convective envelopes become vanishingly thin \citep{Kraft1967}. A piecewise linear fit to the ridge in the CKS sample confirmed that the inflection point occurs at \teff~=~6224~$\pm$~24~K, where the uncertainty was estimated from Markov chain Monte Carlo (MCMC) sampling. This change in slope may be due to the fact that the convective turnover timescale, \taucz, changes rapidly above 6250~K.

In the CKS sample, there also appears to be a clustering of stars above the ridge with \teff~$>6100$~K (seen most clearly in Figure~\ref{fig:inflection}). This cluster of points has a similar slope in the \teff--\prot plane as the long-period pileup, and does not reside on the harmonic of the ridge line as one might expect if the periods were erroneously determined. A similar clustering of points is not observed in the \lamostmcq sample and is less pronounced or absent when substituting the CKS temperatures with \teff from either \citet{Brewer2018} or \citet{Martinez2019}, two studies that independently derived spectroscopic parameters from the CKS spectra. Further inspection of the stars in this cluster revealed that they have anomalously large \teff discrepancies between the CKS and SPOCS catalogs, such that the SPOCS temperatures shift the stars onto the long-period pileup. We conclude that the temperatures for these stars in the CKS catalog are too high by $\gtrsim$~100~K.

A secondary pileup at the short-period edge is also apparent, though less pronounced, in the \teff--\prot distribution of the \lamostmcq sample. We verified through inspection that the secondary pileup does not lie along the \prot/2 harmonic line of the long-period pileup (see Appendix~\ref{app:harmonics}). As shown in the middle panel of Figure~\ref{fig:kde} and the top panel of Figure~\ref{fig:mcmc}, this secondary pileup is seen most clearly through applying Gaussian kernel density estimation (KDE), which was performed with the \texttt{seaborn} package \citep{seaborn}. The short-period pileup is also subtly apparent in the upper right panel of Figure~\ref{fig:xmatch}, as well as scatter plots of the \lamostmcq sample, e.g. the second panel of Figure~\ref{fig:surveys} and the bottom panels of Figures~\ref{fig:models} and \ref{fig:shifted}.

\subsection{Constant Rossby model}
\label{subsec:rossby}
In the WMB model of \citet{vanSaders2016, vanSaders2019}, a star spins down until it reaches a critical Rossby number, at which point magnetic braking ceases. Since Rossby number is highly sensitive to temperature through its dependence on the convective turnover timescale ($\mathrm{Ro} = P_\mathrm{rot}/\tau_\mathrm{cz}$, where $\tau_\mathrm{cz}$ is the convective turnover timescale), this critical threshold corresponds to different rotation periods for stars of different \teff, leading to a pileup in the \teff-\prot plane. Using a small sample of Kepler targets with rotation periods determined from brightness modulations, \citet{vanSaders2016} proposed this threshold happens at a critical Rossby number of $\mathrm{Ro_{crit}} \sim \mathrm{Ro_\odot}$. 

We tested the hypothesis that the long-period pileup observed in the \lamostmcq sample is compatible with the WMB model by fitting constant Rossby models to the \prot boundary. For a given \teff, this model predicts \prot as: 

\begin{equation} \label{eq:1}
    \prot (\mathrm{Ro}, \teff) = \mathrm{Ro} \times \tau_\mathrm{cz}(\teff),
\end{equation}

where we used the equation for the convective turnover timescale (valid in the \teff range of 3300~K$\lesssim \teff \lesssim$~7000~K) presented in \citet{CranmerSaar2011} from a fit to the zero-age main-sequence stellar models of \citet{Gunn1998}:

\begin{multline} \label{eq:2}
\tau_\mathrm{cz}(\teff) = 314.24\exp \left [ -\left (\frac{T_\mathrm{eff}}{1952.5 \mathrm{K}}  \right ) - \left (\frac{T_\mathrm{eff}}{6250 \mathrm{K}}  \right )^{18} \right ] \\+ 0.002.
\end{multline}

The above relation yields a convective turnover timescale for the Sun of $\tau_\mathrm{cz,\odot} = 12.88$~d, leading to a Rossby number of $\mathrm{Ro_\odot}=1.97$ given the mean equatorial rotation period of $P_\mathrm{rot,\odot} = 25.4$~d. For comparison, in the model grids used to create the population models in \S\ref{subsec:models} the solar Rossby number is $\approx$2.16.

\begin{figure}
    \centering
    \includegraphics[width=\linewidth]{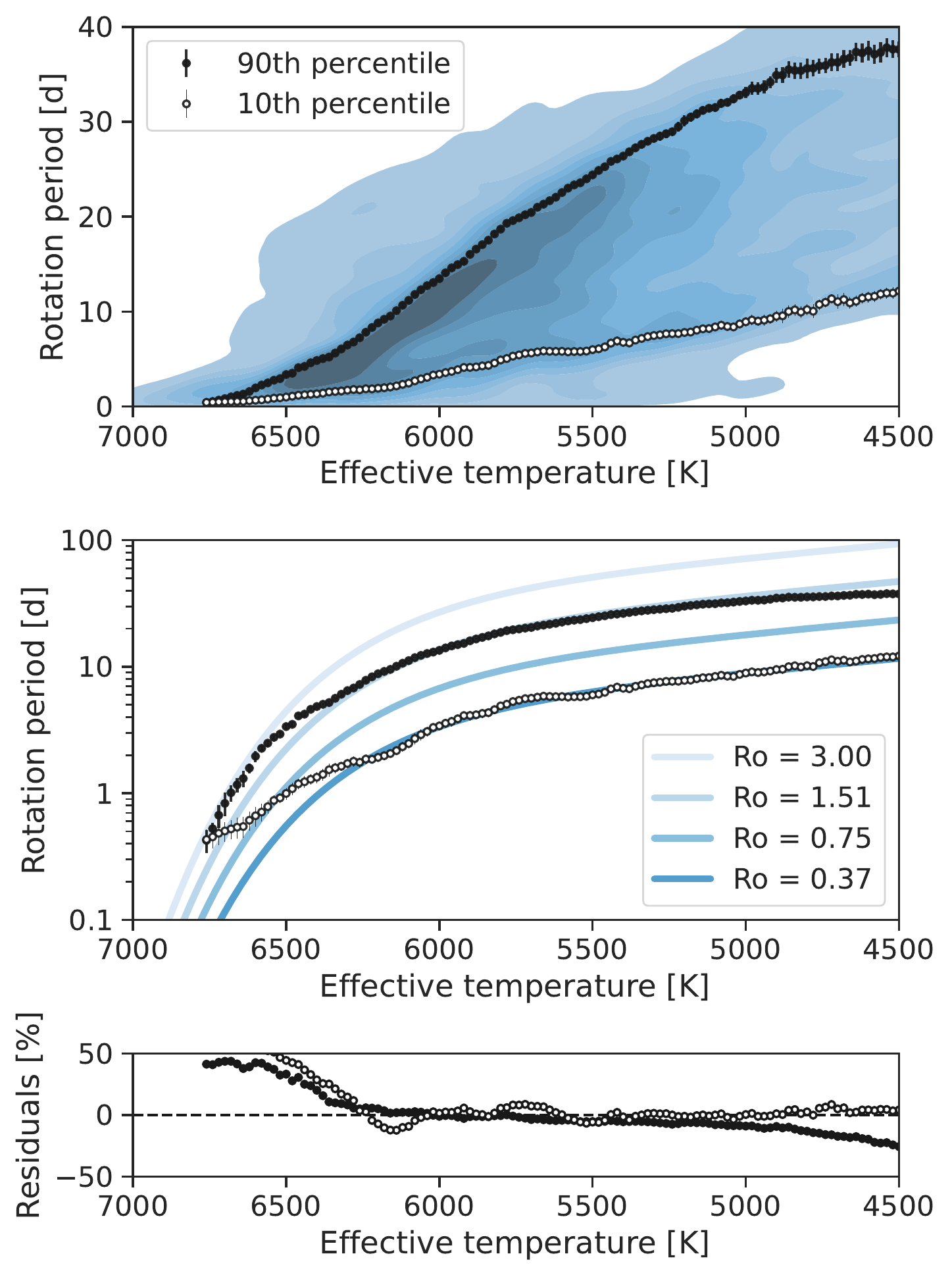}
    \caption{\textit{Top:} Gaussian kernel density estimation of the LAMOST--McQuillan sample and the 10th and 90th \prot percentiles (white and black points, respectively), computed as described in \S\ref{subsec:rossby}. \textit{Middle:} The same 10th and 90th percentile curves shown in relation to constant Rossby curves. \textit{Bottom:} Residuals from the median models resulting from the MCMC sampling. The fits depicted above do not include a \teff offset.}
    \label{fig:mcmc}
    \script{mcmc.py}
\end{figure}

We approximated the long-period edge in the following manner. For 100 bootstrapped resamplings of the \teff–\prot distribution, leaving out 50\% of the data for each bootstrapped sample, we computed the 90th percentile of \prot values in overlapping \teff bins with centers located every 20~K between 4000~K and 7000~K and half-widths of 100~K. The final long-period edge curve was then computed as the mean of the bootstrapped 90th percentile values. We found it was also necessary to omit stars with Ro~$>$~5/3 in this computation to match the long-period edge. We computed the 10th percentile curve and its uncertainty similarly, as an approximation to the lower boundary of the \teff-\prot plane. We show these curves in relation to the full \lamostmcq sample and to constant Rossby models in Figure~\ref{fig:mcmc}.

We performed an initial Levenberg-Marquardt non-linear least-squares fit of a constant Rossby model to the long-period edge with the \texttt{curve\_fit} function in the \texttt{scipy.optimize} class to optimize the following likelihood: 

\begin{multline} \label{eq:3}
    \ln{p} (y | T_\mathrm{eff}, T_\mathrm{sys}, \sigma, \mathrm{Ro}, f) =\\ -\frac{1}{2}\sum_n \left [ \frac{(y_n - P_\mathrm{rot}(\mathrm{Ro}, T_\mathrm{eff} + T_\mathrm{sys}))^2}{s_n^2}  + \ln{(2\pi s_n^2)} \right ],
\end{multline}

where

\begin{equation} \label{eq:4}
    s_n^2 = \sigma^2 + f^2 P_\mathrm{rot}(\mathrm{Ro}, T_\mathrm{eff} + T_\mathrm{sys})^2,
\end{equation}

and $y_n$ is the value of a \prot percentile curve in the $n$th \teff bin. This is a Gaussian likelihood where the variance is underestimated by some fraction, $f$. Here $T_\mathrm{sys}$ is a constant to allow for a systematic offset between the data and the models used to calibrate the \taucz relation. This offset can equivalently be thought of as a correction to the data, or a correction to the model \teff scale. We performed Markov chain Monte Carlo sampling (MCMC) of this likelihood with the \texttt{emcee} package \citep{emcee2013, emcee2019} to estimate the mean and uncertainty of the critical Rossby number that best matches the long-period edge in the range of 5000~K$\lesssim \teff \lesssim$~6250~K. We instantiated 32 walkers around the least-squares solution and sampled for $10^5$ steps, adopting uniform priors on Ro, $f$, and $T_\mathrm{sys}$ with ranges of (0.1,10), (0,10), and (-1000~K, 1000~K), respectively. Convergence was assessed by ensuring the chain length was at least 100 times longer than the chain autocorrelation lengths for each parameter. A similar analysis was performed for the 10th percentile curve, restricted in the range of 4500~K$\lesssim \teff \lesssim$~5800~K where a constant Rossby model provides a reasonable fit. 

We additionally fit the CKS long-period pileup (using three different homogeneous \teff sources) and the \hall asteroseismic main-sequence sample, allowing for \teff offsets in each data set (Table~\ref{tab:mcmc}). For the CKS pileup stars and the \lamostmcq percentile curves, we assumed constant fractional \prot uncertainties of 10\%. To isolate the long-period pileup stars in the CKS sample, we selected a trapezoidal region using the condition $-0.0314~\teff + 199.286 < \prot < -0.0314~\teff + 206.286$ (shown in Figure~\ref{fig:inflection}). We note that this selection is particular to the CKS \teff scale and is not general. For the CKS sample, we additionally required $\logg~>~4$ and 5850~K~$<~\teff~<$~6500~K. 

The constant Rossby model provides a reasonably good description of the \lamostmcq long-period edge in the \teff range of $\approx$~5000--6250~K, with fractional residuals $\lesssim5\%$ over this range. Above and below this \teff range we see clear and significant divergence from the constant Rossby model, such that the model under-predicts periods of hotter stars and over-predicts periods of cooler stars, possibly because the cooler stars have not had enough time to evolve to the critical Rossby number associated with weakened magnetic braking \citep[see Figure 6 in][]{vanSaders2019}.

In some cases, our constant Rossby fits to the long-period pileup prefer Rossby numbers lower than the Sun's, in contrast with \jvs. We caution against overinterpreting the specific \rocrit values inferred here, and discuss our interpretations further in \S\ref{subsec:longperiod}.

\begin{figure*}
    \centering
    \includegraphics[width=\linewidth]{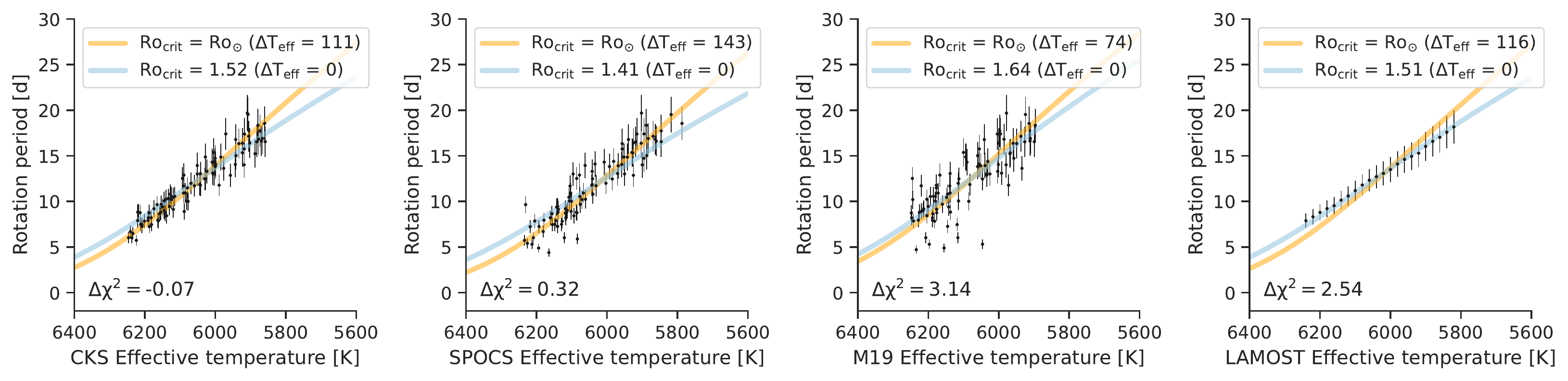}
    \caption{Non-linear least-squares fits to the long-period pileup for a \rocrit~=~\rosun model with constant \teff offset (orange) and a variable \rocrit model with no \teff offset (blue). The $\Delta \chi^2$ values printed in each panel are computed as $\chi^2_{\text{Ro}_\text{crit}=\text{Ro}_\odot} - \chi^2_{\Delta T_{eff}=0}$.}
    \label{fig:rocrit}
    \script{rocrit.py}
\end{figure*}

To further quantify the preference for \rocrit~$<$~\rosun we performed non-linear least-squares fits of two models to the CKS long-period pileup and the \lamostmcq 90th percentile curve (both in the \teff range of 5800--6250~K). The first model assumes \rocrit = \rosun and has one free parameter, a temperature offset added to the data. The second model assumes no temperature offset and allows \rocrit to vary (\rocrit being the only free parameter). The results of these fits are shown in Figure~\ref{fig:rocrit}. We found that for the CKS sample, particularly when using the more precise temperatures from \citet{Fulton2018}, that the $\Delta \chi^2$ between the \rocrit = \rosun and variable \rocrit models was negligible. In other words, the \rocrit = \rosun model fits the data well with a \teff shift of $\approx$~70--140~K depending on the \teff source used. For the \lamostmcq sample, there is weak support for the variable \rocrit model. However, as shown in Appendix~\ref{app:teff}, there is a strong systematic trend not adequately captured by a constant offset (i.e. a \teff-dependent \teff offset) in the LAMOST temperatures when compared to other spectroscopic catalogs (particularly SPOCS). Correcting for these systematic trends weakens the support for the variable \rocrit model, and we speculate that temperature systematics bias the \rocrit inference procedure. 

\begin{deluxetable*}{lllll}
\tabletypesize{\scriptsize}
\tablecolumns{5}
\tablewidth{0pt}
\tablecaption{Results of constant Rossby model fits.\label{tab:mcmc}}
\tablehead{
\colhead{Sample} & 
\colhead{\teff range} & 
\colhead{Ro} & 
\colhead{f} & 
\colhead{$T_\mathrm{sys}$ (K)} 
}
\startdata
\textit{Short-period pileup} \\
\hline
\lamostmcq 10th \prot pctl. & 4500~K--5800~K & 0.37 $\pm$ 0.02 & 0.011 $\pm$ 0.009 &  -46 $\pm$ 80 \\
\hline 
\textit{Long-period pileup} \\
\hline
\lamostmcq 90th \prot pctl. & 5800~K--6250~K & 1.4 $\pm$ 0.1 & 0.019 $\pm$ 0.015 & -44 $\pm$ 43 \\
CKS (CKS \teff) & 5800~K--6250~K & 1.77 $\pm$ 0.07 & 0.02 $\pm$ 0.01 & 70 $\pm$ 15 \\ 
CKS (SPOCS \teff) & 5800~K--6250~K & 1.9 $\pm$ 0.1 & 0.12 $\pm$ 0.01 & 136 $\pm$ 23 \\
CKS (M19 \teff) & 5800~K--6250~K & 1.6 $\pm$ 0.2 & 0.18 $\pm$ 0.02 & -21 $\pm$ 42 \\
\citet{Hall2021} main-sequence & 5800~K--6250~K & 1.9 $\pm$ 0.4 & 0.27 $\pm$ 0.05 & 29 $\pm$ 79 \\
\enddata
\end{deluxetable*}

\subsection{Comparison with theory}
\label{subsec:models}

We compared the \teff--\prot distribution of the CKS and \lamostmcq samples with the theoretical predictions of \citet{vanSaders2019}, which were updated for \citet{Hall2021} and utilized in that work. Those authors presented forward modeling simulations of the observed Kepler \prot distribution including theoretical models of stellar angular momentum evolution (for both the standard spin-down and WMB scenarios), a galactic population model, and a prescribed observational selection function. For this exercise we selected main-sequence stars from the simulations using an evolutionary state flag, defining main-sequence stars as those with a He core mass fraction $<$0.0002.\footnote{These models are packaged in a Zenodo repository at \url{https://doi.org/10.5281/zenodo.6471539}.} We did not apply a cut on Rossby number to the models to mimic a detection threshold, as one might expect if detectability is Ro-dependent. The issue of detection bias is discussed further in \S\ref{subsec:detectionbias}.

\begin{figure*}
    \centering 
    \includegraphics[width=0.49\linewidth]{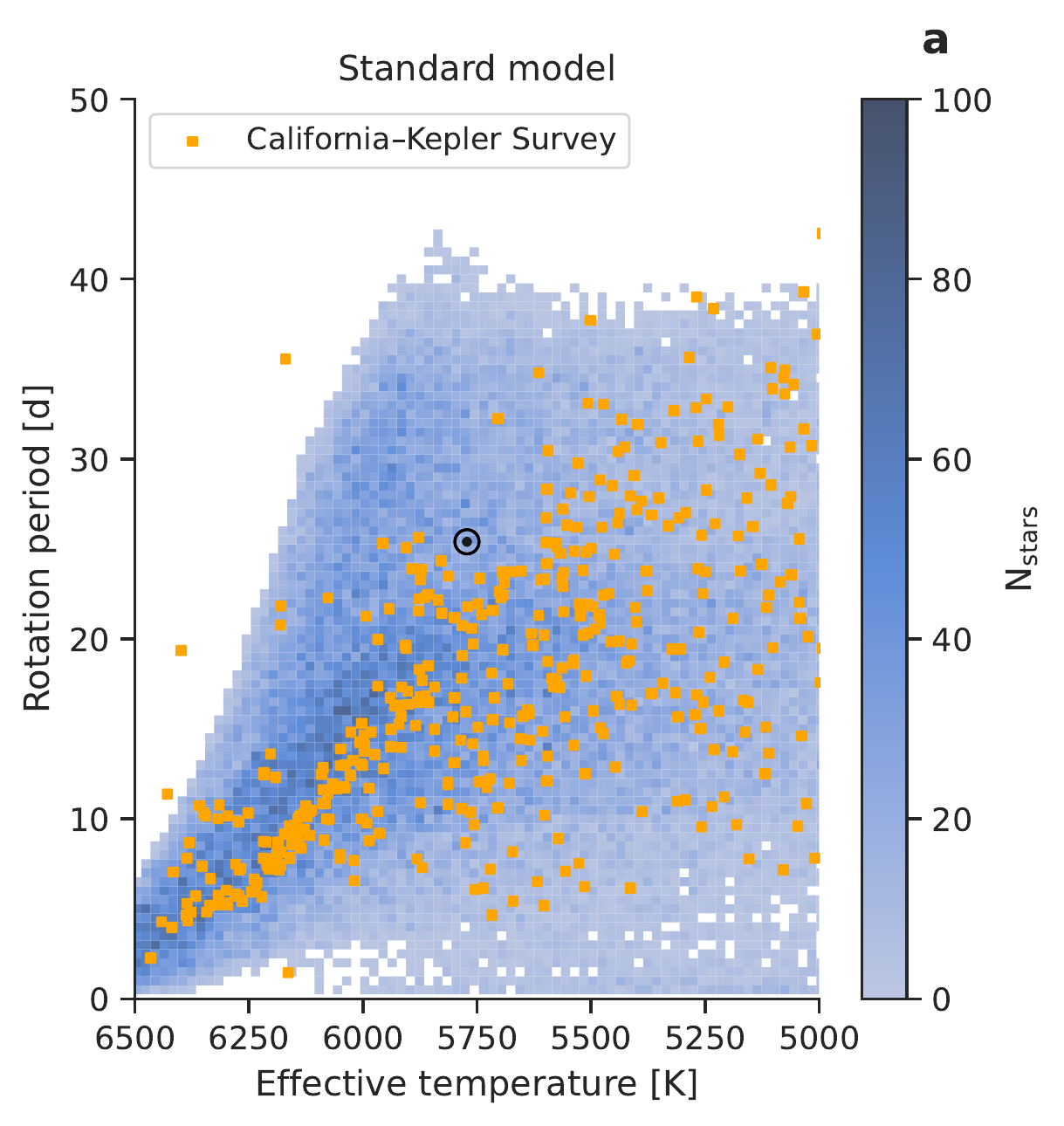}
    \includegraphics[width=0.49\linewidth]{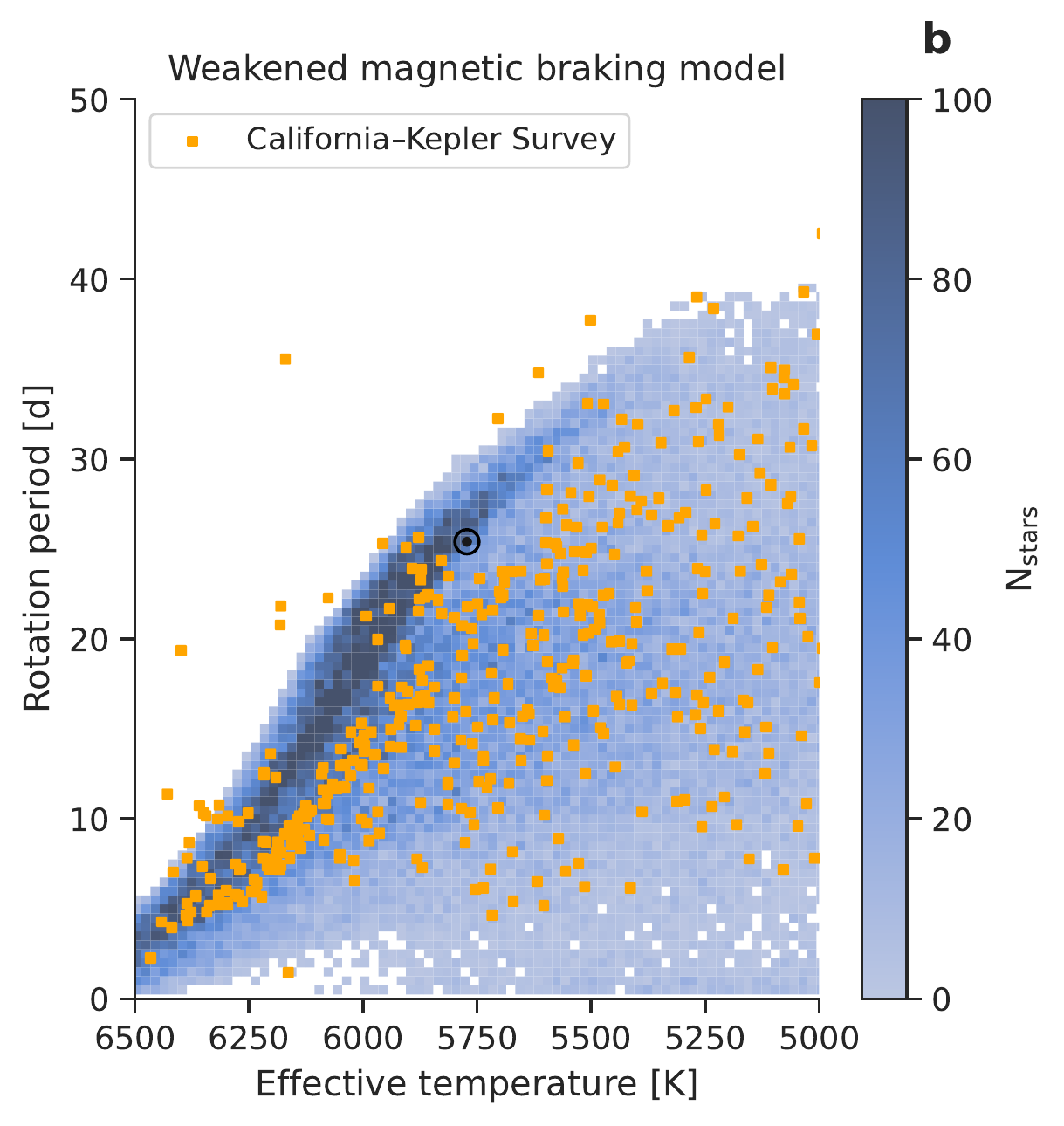}
    \includegraphics[width=0.49\linewidth]{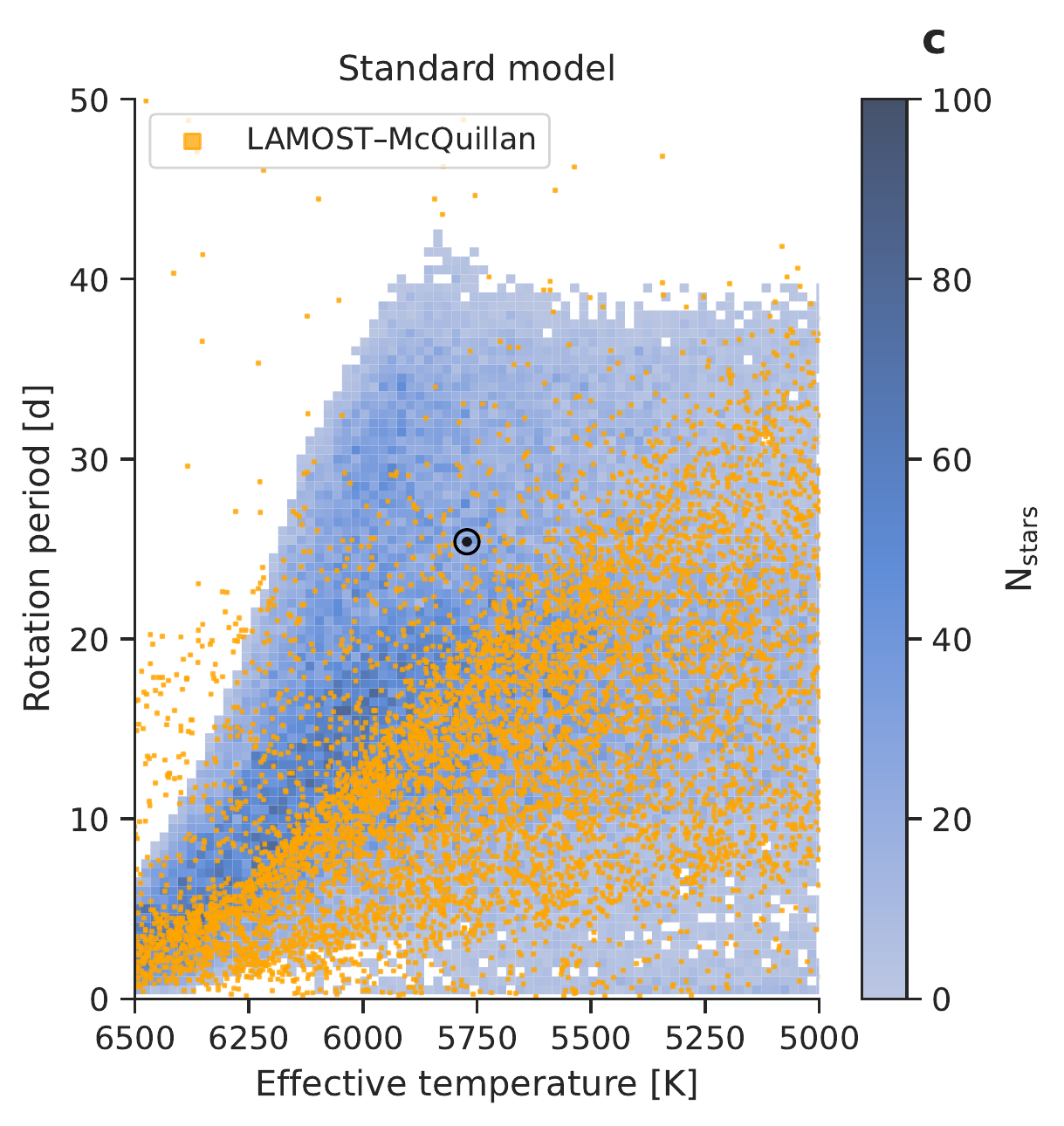}
    \includegraphics[width=0.49\linewidth]{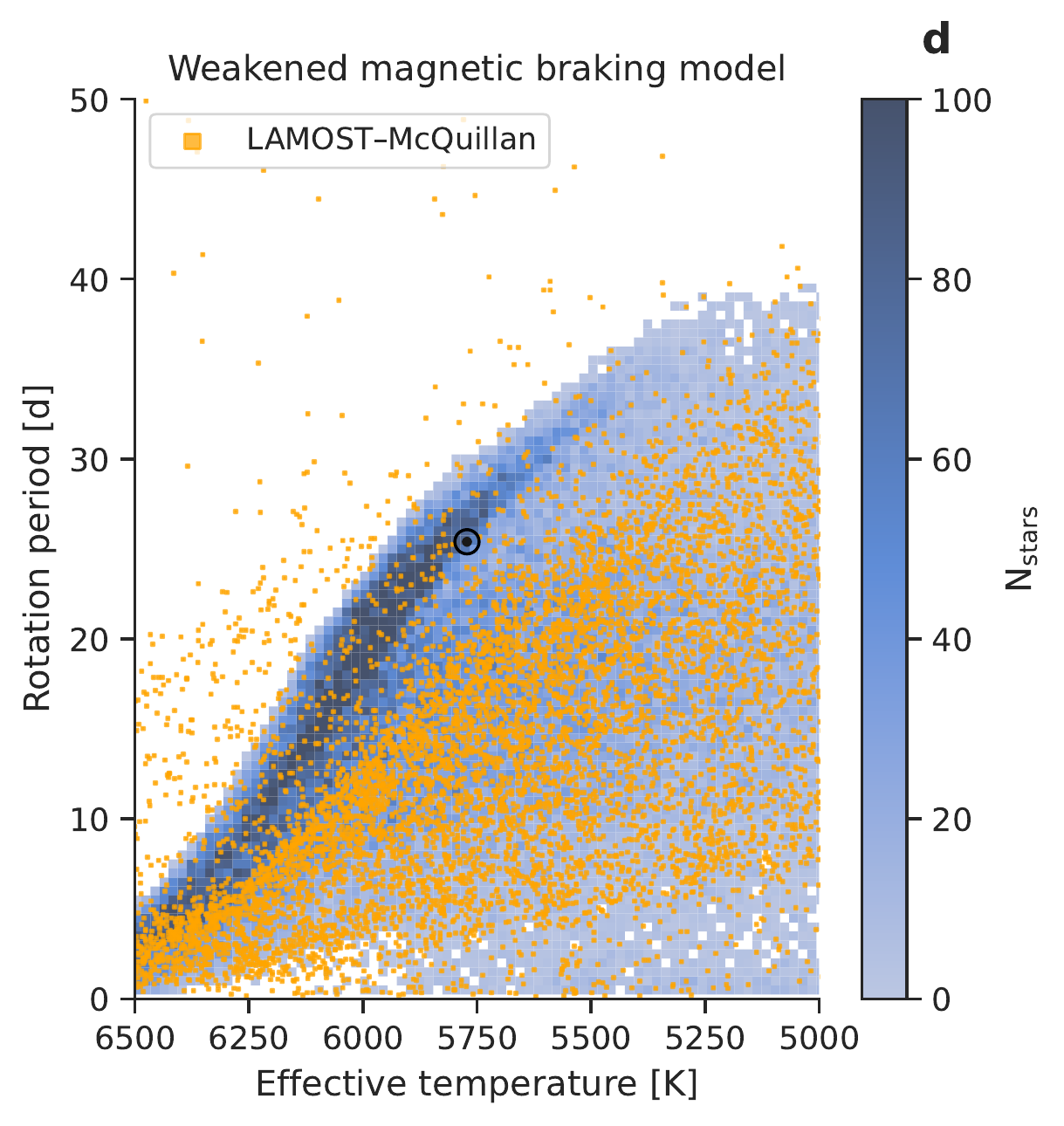}
    \caption{The \teff-\prot plane for the CKS sample (top panels) and LAMOST sample (bottom panels) in comparison to the standard and WMB models (2-d histograms) presented in \jvs. Shown here are stars with \logg~$>$~4.1. Rotation periods for the CKS sample here are sourced from the \citet{David2021} compilation. The black symbol in each panel indicates the position of the Sun.}
    \label{fig:models}
    \script{models.py}
\end{figure*}

\begin{figure*}
    \centering 
    \includegraphics[width=0.49\linewidth]{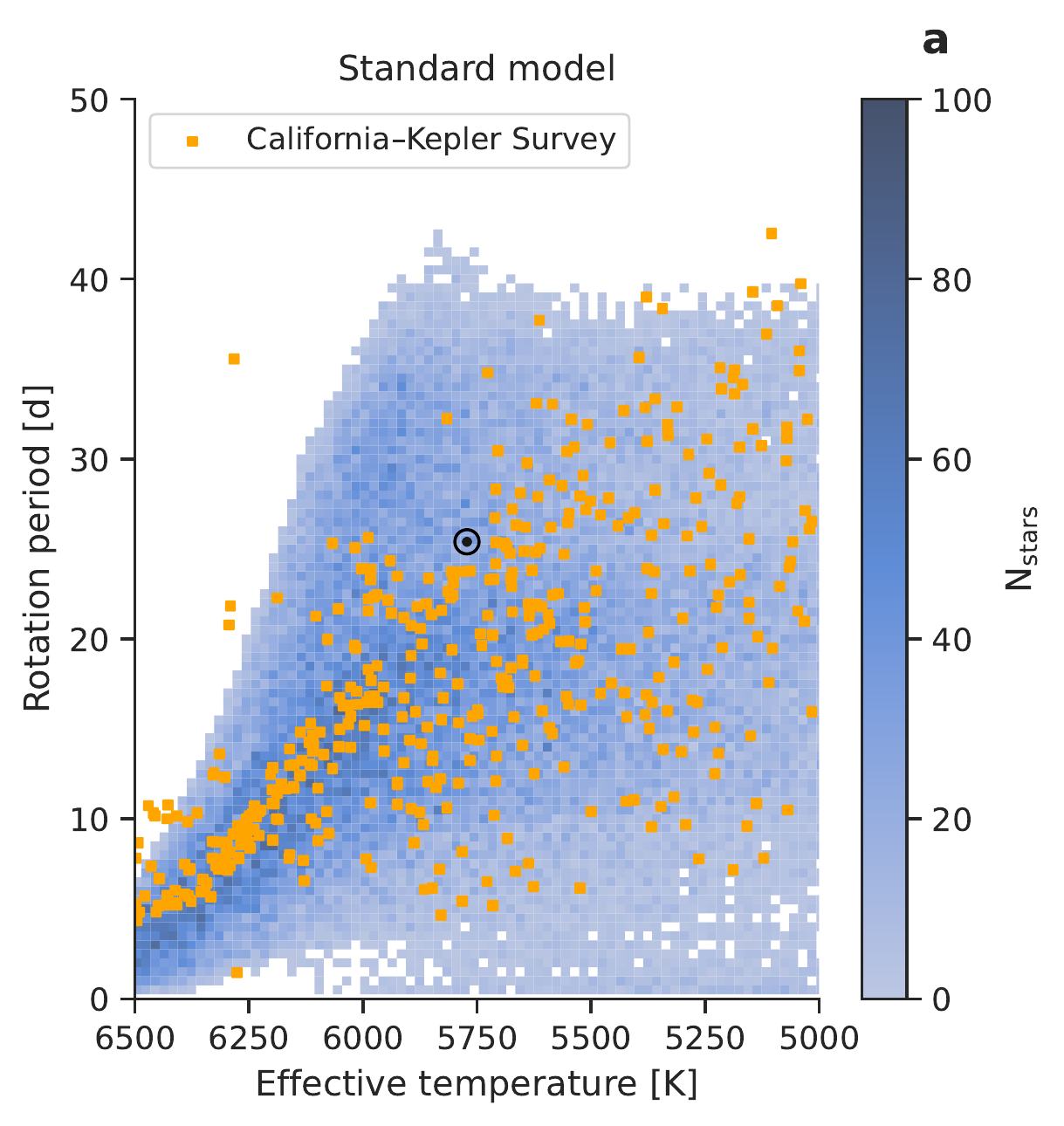}
    \includegraphics[width=0.49\linewidth]{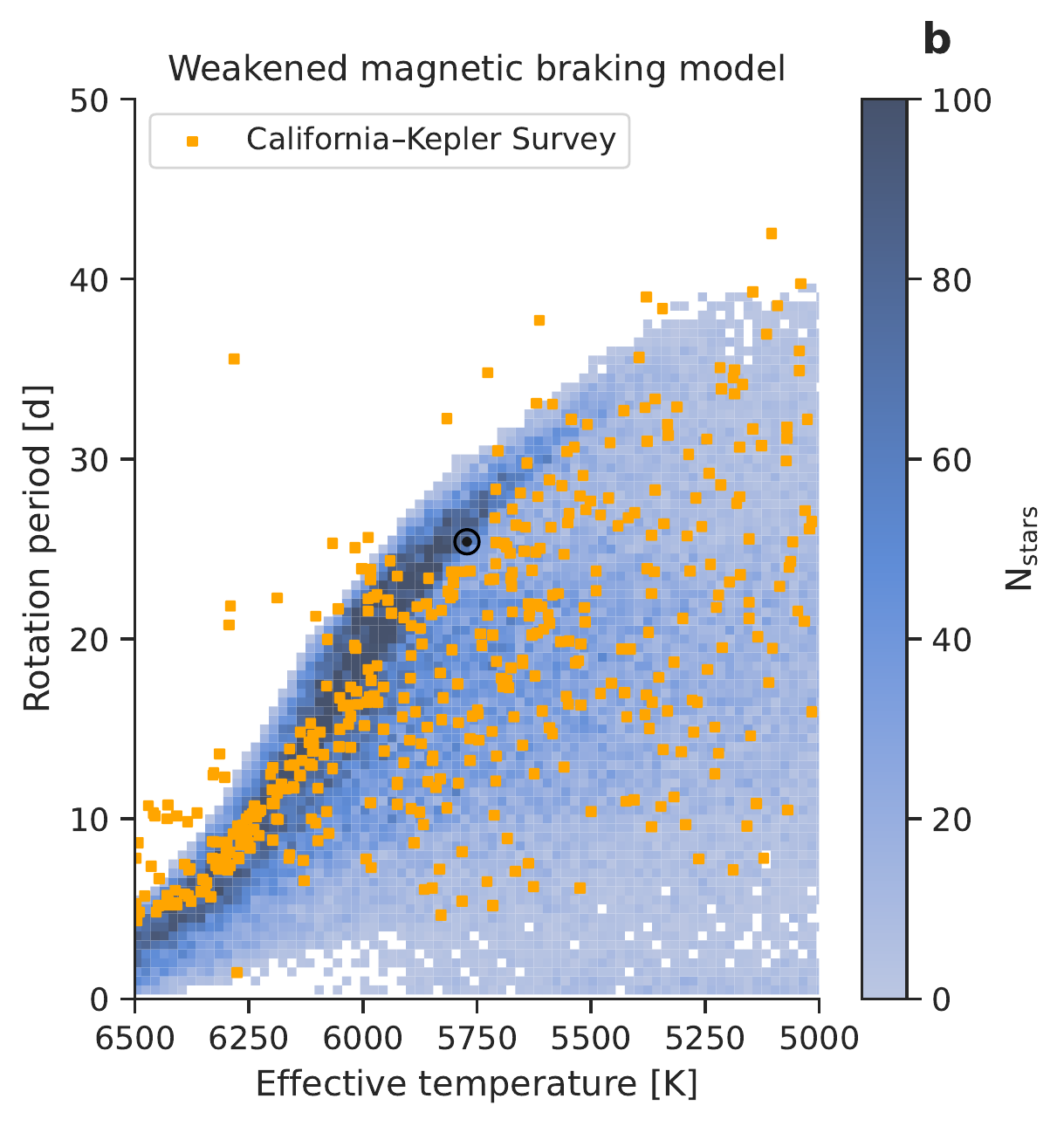}
    \includegraphics[width=0.49\linewidth]{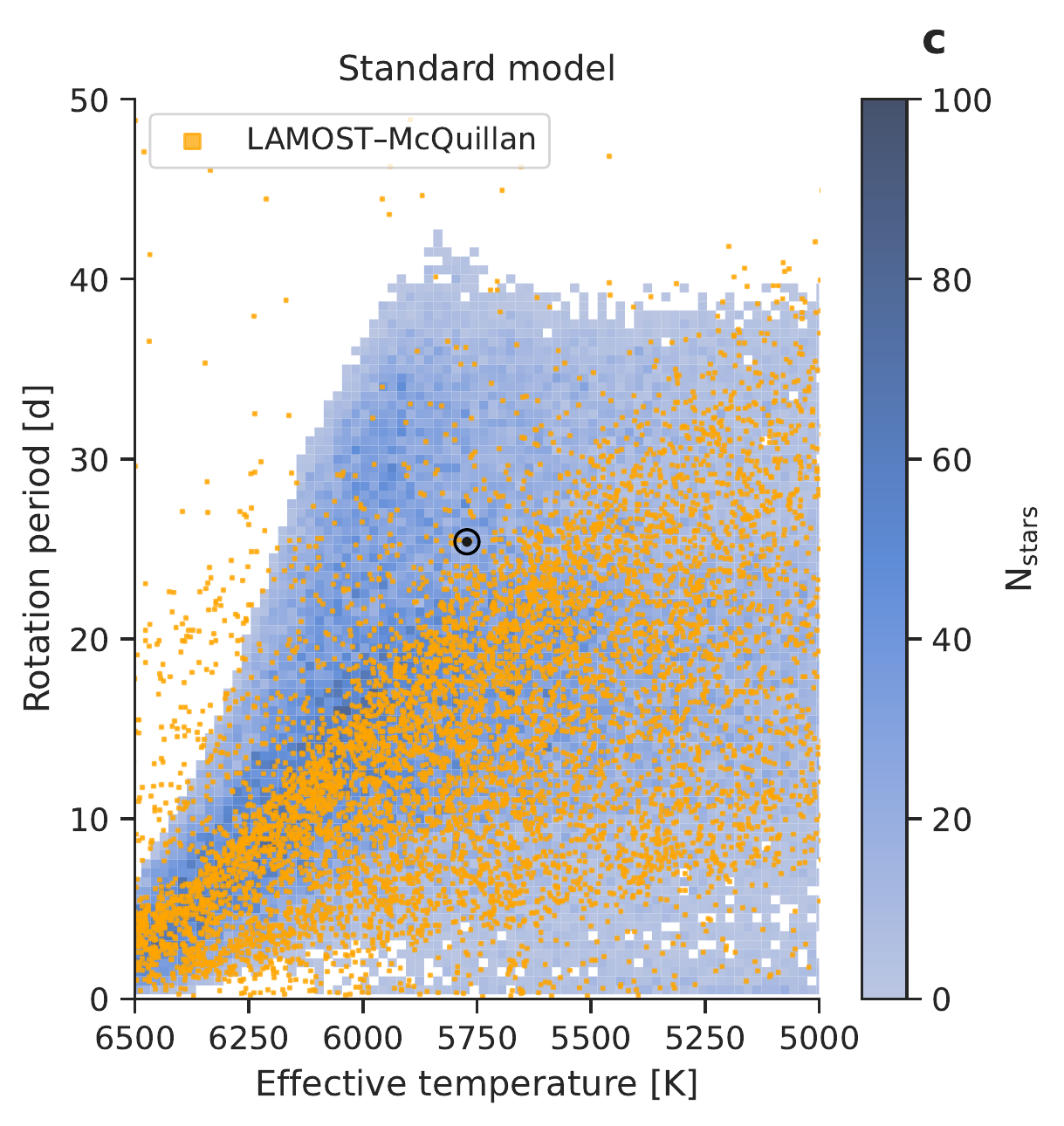}
    \includegraphics[width=0.49\linewidth]{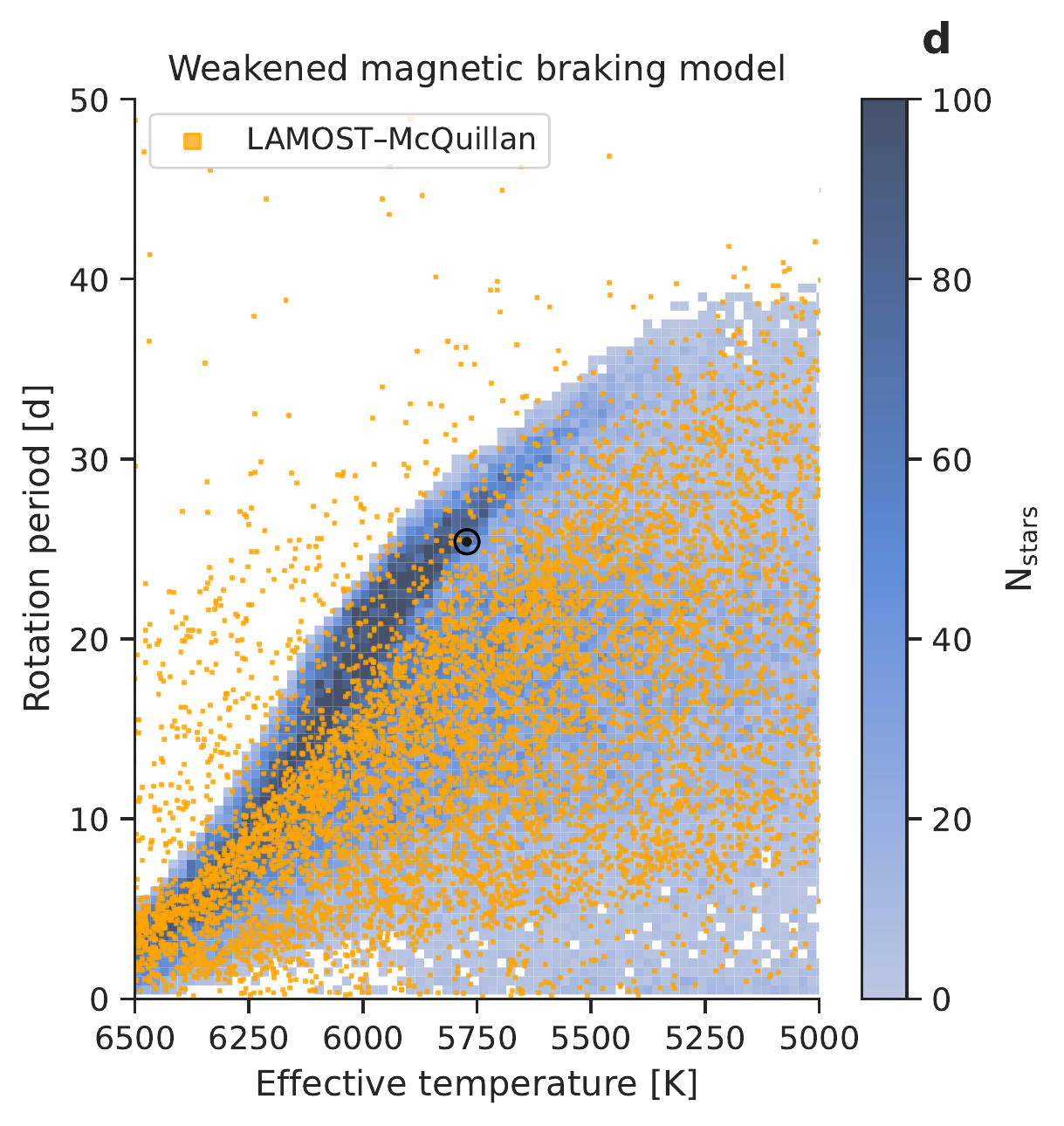}
    \caption{Same as Figure~\ref{fig:models} with constant \teff offsets applied to the data. Shifts of +111~K and +116~K are applied to the CKS and LAMOST \teff, respectively. The \teff shifts originate from a least-squares fit of a \rocrit~=~\rosun curve (with a \teff offset) to the long-period pileup (as described in \S\ref{subsec:rossby}).}
    \label{fig:shifted}
    \script{shifted.py}
\end{figure*}

The theoretically predicted \teff--\prot distributions of \jvs are shown in relation to the observations in Figure~\ref{fig:models}. Neither model satisfactorily matches the observations, though the WMB model more closely matches the long-period edge of F-type and early G-type stars. The specific WMB prescription of \jvs adopted a critical Rossby number of \rocrit~=~2.08 (using a different \taucz prescription than the one used here), leading to a pileup that is located at larger \prot values (at fixed \teff) when compared to the observations, assuming no \teff offset between the models and the data. Figure~\ref{fig:shifted} shows the same models in relation to the data with constant \teff offsets applied to the data, which are derived in \S\ref{subsec:rossby}. The \teff offsets lead to better agreement between the data and models, although the long-period pileup in the \lamostmcq sample appears to overlay the models only for \teff~$\gtrsim 6000$~K, possibly due to strong \teff-dependent systematics in the LAMOST \teff scale (see Appendix~\ref{app:teff}).

\begin{figure}
    \centering
    \includegraphics[width=\linewidth]{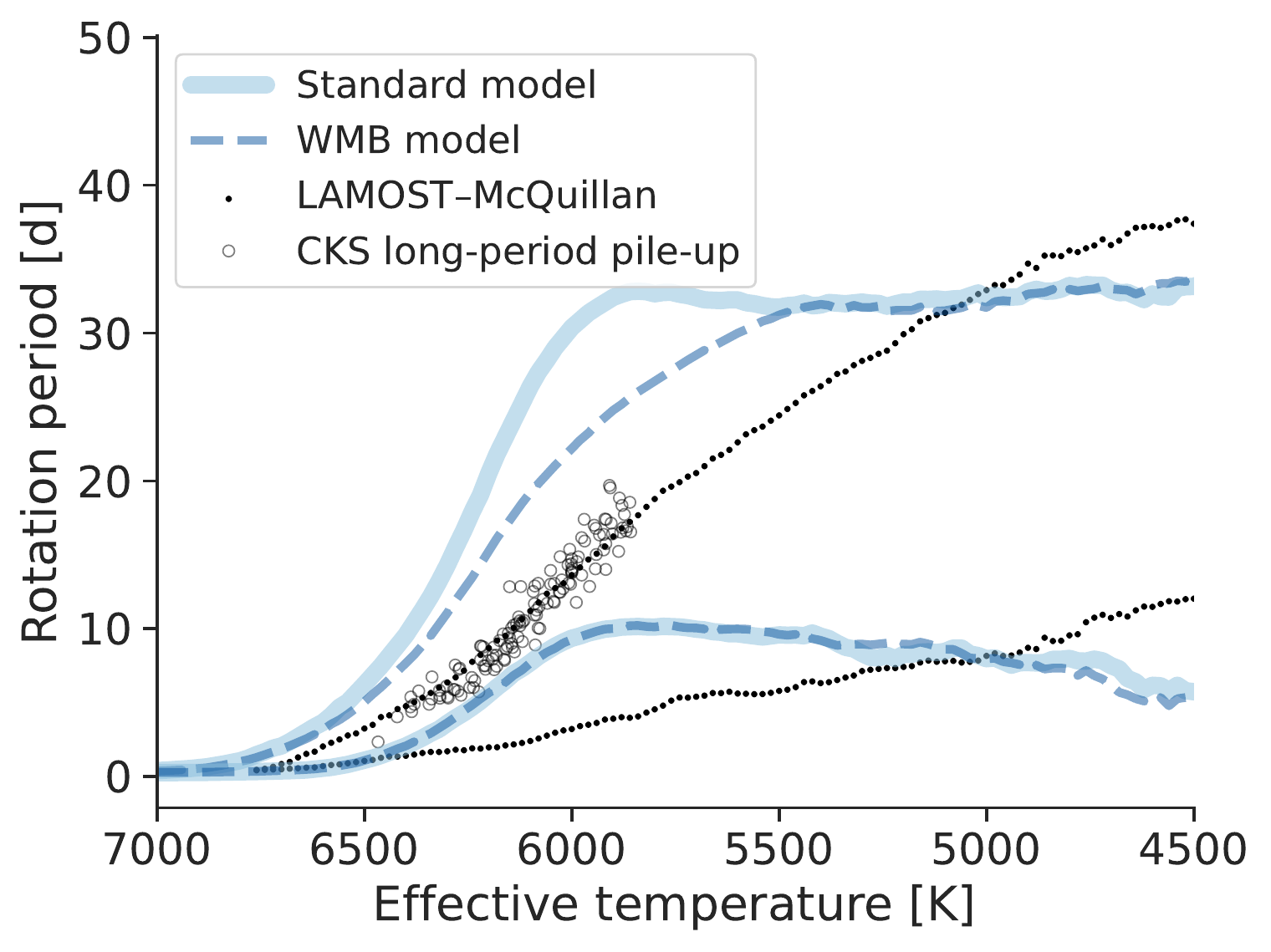}
    \caption{Comparison of the theoretically predicted \teff--\prot distribution of Kepler stars from \jvs (given by the 10th and 90th \prot percentiles) with the same percentile ranges from the \lamostmcq sample (black points) and the CKS long-period pileup (open circles).}
    \label{fig:percentiles}
    \script{percentiles.py}
\end{figure}

To quantify the degree of agreement between the theoretical models and the \lamostmcq observations we computed the 10th and 90th percentile \prot ranges of the standard and WMB models in overlapping \teff bins, analogous to how the upper and lower boundaries of the observed \prot distribution were found in \S\ref{subsec:rossby}. We computed the $\chi^2$ values between the observed upper edge and the 90th percentile ranges of the standard and WMB models, finding the WMB model is preferred with a $\Delta \chi^2 = 154$. Moreover, the WMB model better reproduces the slope of the observed long-period edge between 5300--6000~K (Figure~\ref{fig:percentiles}). 

While better agreement between the WMB model and observations might be achieved with stalling at a lower \rocrit, it is also possible that there are systematic offsets in the \teff scales between the observations and models used in \jvs, as well as differences in the computation of \taucz.  We also note that, while the models were computed using a simulated Kepler stellar population and selection function, the actual observed population and selection function of the \lamostmcq may be slightly different. 

Shifting the LAMOST \teff to higher values would bring the data into better agreement with the models, and in Appendix~\ref{app:teff} we show the LAMOST \teff are $\sim$50--100~K cooler than the other surveys considered here. In turn, it appears that the long-period edge for lower mass stars would be at higher \prot than the models (i.e. the low-mass stars would be rotating more slowly than the model predictions). Such a discrepancy could result from different underlying populations between the models and the \lamostmcq sample, or a different normalization for the magnetic braking law. The models above employ a modified magnetic braking law that is scaled to match the rotation period of the Sun \citep[see equations 1 \& 2 of][]{vanSaders2013}, with a normalization factor of $f_K = 6.6$. A higher normalization factor would cause the low-mass stars to spin down more at fixed age.

Notably, both models fail to reproduce the observed short-period pileup, possibly due to the assumption of solid-body rotation in both models. At early times, Sun-like and low-mass stars are expected to have strong radial differential rotation due to their rapid collapse onto the main-sequence. The core and envelope at these times are thus assumed to be decoupled. However, the core and envelope are expected to couple on timescales of a few tens of million years for Sun-like stars \citep{Denissenkov2010, GalletBouvier2015, Lanzafame2015} or hundreds of million years for low-mass stars \citep{GalletBouvier2015,Lanzafame2015,Somers2016}. When this happens, angular momentum can be transferred from the core to the envelope at a rate comparable to the rate at which angular momentum is lost via magnetized winds. Consequently, so-called ``two-zone'' models \citep{MacGregor1991} spin more rapidly than solid-body rotators at the same age. This hypothesis is discussed further in \S\ref{subsec:shortperiod}.

\subsection{Comparison with literature surveys}
\label{subsec:asteroseismic}

\hall determined rotation periods for 91 main-sequence, asteroseismic Kepler targets from rotational splitting of asteroseismic oscillation frequencies. We found that the distribution of the asteroseismic sample in the \teff--\prot plane approximately matches the pileup we observe, although the \hall sample appears shifted slightly to either higher \prot or higher \teff values relative to the ridge in the \lamostmcq sample while such an offset is either absent or not as apparent relative to the CKS sample. Such an offset could be due to differing spectroscopic temperature scales between the two studies, differing observational biases, or some combination of effects.

To assess the consistency of the \teff scales between catalogs, we cross-matched the \hall sample with the LAMOST DR5 catalog and the CKS sample. We found a root mean square (RMS) of the residuals between the \hall and LAMOST DR5 \teff measurements of 55~K, with a median offset of 41~K, such that the LAMOST \teff scale is cooler, on average (see Appendix~\ref{app:teff}). Similarly, we found excellent agreement between the \teff scales of the CKS and \hall samples, with a median offset of 29~K (such that the CKS scale is hotter) and an RMS of 41~K. These offsets are modest, and do not account for non-linear systematics which may exist (particularly in the LAMOST \teff).

\masuda inferred the \prot distribution of 144 late-F/early-G dwarfs in the Kepler field from precise stellar radii and spectroscopically determined \vsini. As with the asteroseismic sample, we compared the CKS and \lamostmcq samples with the results of those authors. We find that the long-period pileup we observe appears to overlap with the \masuda data, within the uncertainties of those authors. We therefore consider it likely that the long-period pileup is the same feature, or an extension of the same feature, that was detected in the \hall and \masuda studies.

\begin{figure*}
    \centering
    \includegraphics[width=\linewidth]{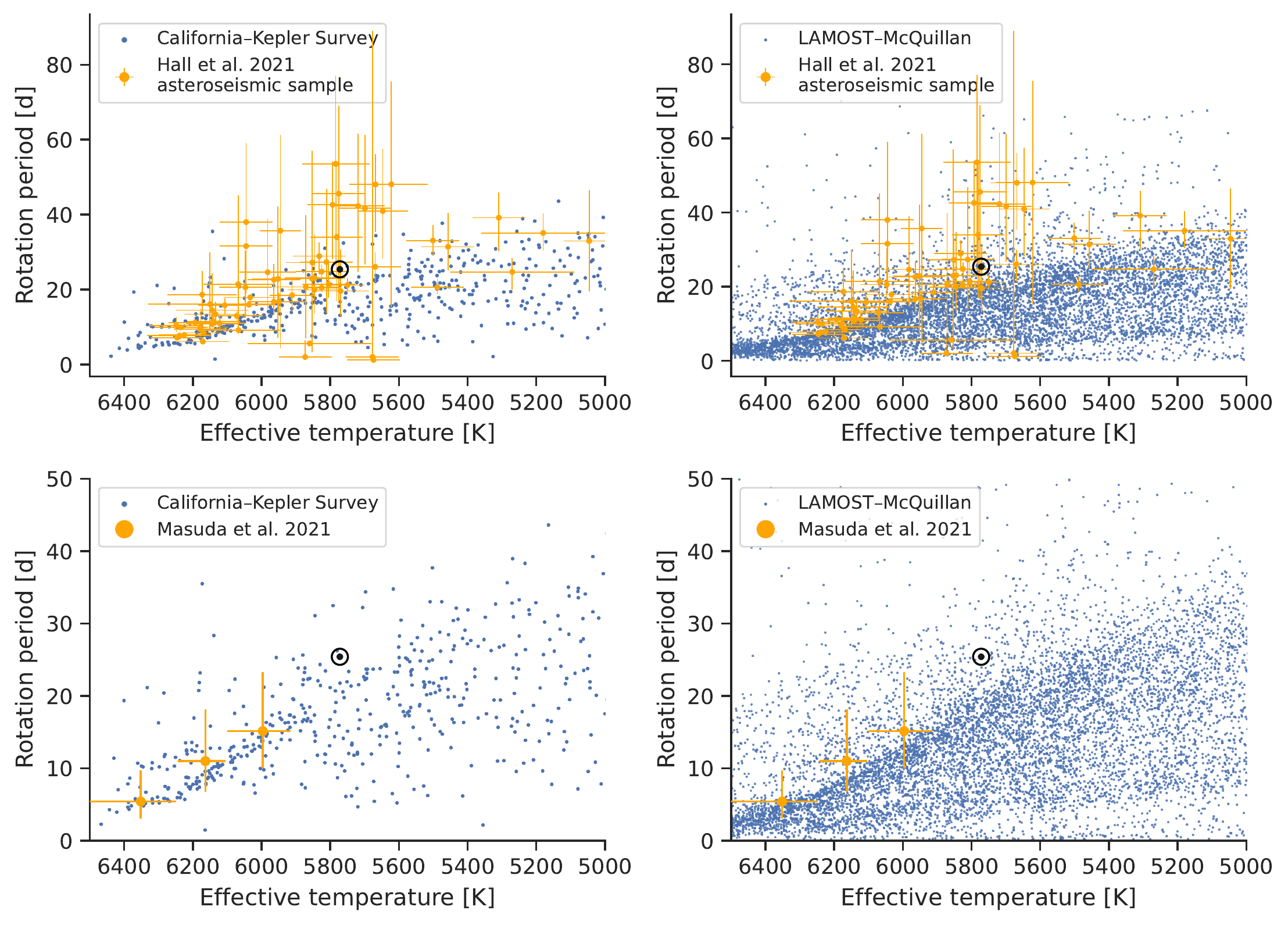}
    \caption{Comparison of CKS (left column) and \lamostmcq samples (right column) with the \hall main-sequence asteroseismic sample (top row) and the \masuda sample (bottom row) in the \teff--\prot plane. The black point indicates the Sun. Note, the CKS, LAMOST, and \hall samples derive \teff from distinct pipelines. In the top panels, constant offsets of -29~K and +41~K were added to the CKS and LAMOST \teff, respectively, based on comparisons to those stars with overlap in the H21 sample (see \S~\ref{subsec:asteroseismic} and Appendix~\ref{app:teff}). Using LAMOST \teff, where it exists, for the \hall sample brings that sample into even closer agreement with the long-period pileup in the LAMOST-McQuillan sample.}
    \label{fig:asteroseismic}
    \script{asteroseismic.py}
\end{figure*}

\subsection{Ages of stars on the long-period pileup}
\label{subsec:ages}
\begin{figure*}
    \centering
    \includegraphics[width=\linewidth]{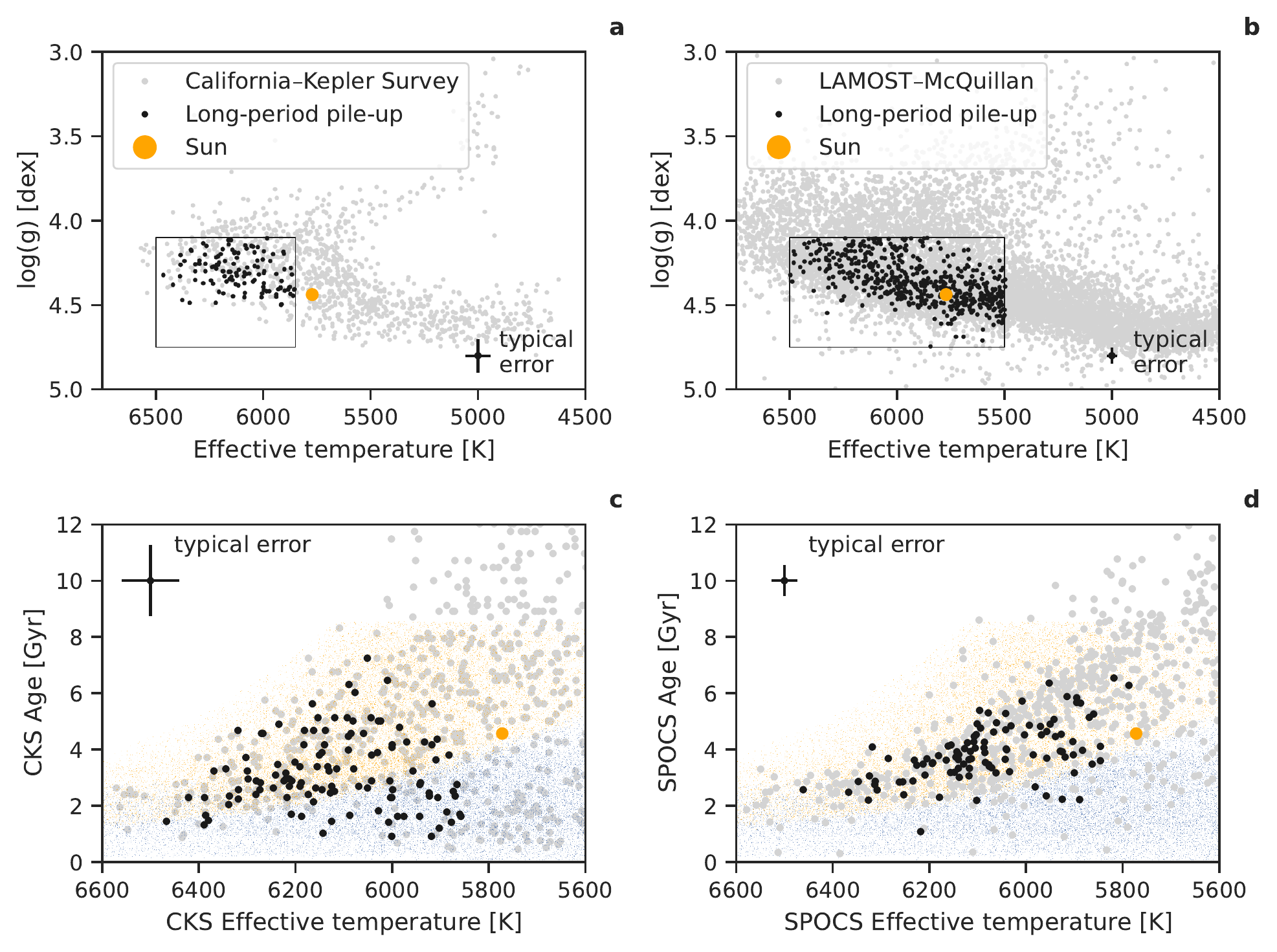}
    \caption{Above, H-R diagram placement of long-period pileup stars relative to the Sun and the CKS sample (a) and similarly for the \lamostmcq sample (b). Below, the \teff-age plane for CKS stars along the long-period pileup using isochrone ages from the CKS (c) and SPOCS (d) catalogs. The light shaded points represent predictions from the WMB model of \jvs for stars with Ro~$<$~2 (blue) and Ro~$>$~2 (orange). Note, panels (a), (c), and (d) all pertain to the CKS sample, whereas panel (b) relates to the \lamostmcq sample, for which homogeneous ages have not been published.}
    \label{fig:ages}
    \script{ages.py}
\end{figure*}

The WMB model predicts that hotter stars pile up on the long-period edge at younger ages, producing an age gradient across the edge. In Figure~\ref{fig:ages} we show the \teff-age distributions of stars on the ridge using ages from the CKS \citep{Fulton2018} and SPOCS \citep{Brewer2018} catalogs, where we use the trapezoidal selection described in \S\ref{subsec:rossby} to select stars on the long-period pileup. Both catalogs derive spectroscopic parameters from the CKS spectra, but use different pipelines for both the spectroscopic parameters and the isochrone fitting. 

In both cases, there appears to be an age gradient such that hotter stars are younger on average. However, such a trend is also expected in a sample of main-sequence stars as a natural consequence of the shorter main-sequence lifetimes of hotter, more massive stars. We also find that the dispersion in age is a sensitive function of \teff, with cooler stars on the ridge showing a broader range of ages. This observation could be due to cooler stars populating the ridge for longer periods of time (relative to hotter stars), the lower precision of isochrone ages for cooler stars, or some combination of the two effects.

We determined that 90\% of the stars on the ridge have ages between 1.4--5.6~Gyr (using ages from the CKS catalog), or 2.3--5.9~Gyr (using SPOCS ages). These ranges are consistent with the range of ages observed in the \hall and \masuda samples, as shown in Figure~11 of the latter reference. However, we note that systematic effects in surveys and theoretical models lead to large uncertainties in isochrone ages that are not necessarily represented by the reported age uncertainties. For example, it is curious that, when using SPOCS ages only $\sim$1\% of the stars have ages $<$~2~Gyr and that stars are concentrated at the upper age boundary for a given \teff (Figure~\ref{fig:ages}). In fact, there are only a handful of long-period pileup stars in the CKS sample with ages older than the age of the Sun (using CKS ages), and almost all of the pileup stars would be compatible with an age equal to or less than the Sun's given the large age uncertainties. 

In the WMB model, although wind-driven angular momentum losses cease, stars continue to evolve structurally which results in evolution in the moment of inertia and stellar spin, driven by expansion of the stars away from the main-sequence \citep{vanSaders2019}. Stars reach \rocrit~$\approx$~2 approximately halfway through their main-sequence lifetimes and remain there until the main-sequence turnoff. Thus, higher mass stars with shorter main-sequence lifetimes should show a smaller age spread on the long-period pileup relative to lower mass stars with longer main-sequence lifetimes. Though we have not quantified such an age-gradient, the data suggest that some stars spend several Gyr occupying the ridge with only modest evolution of their spin rates.

We additionally selected stars on the long-period pileup from the \lamostmcq sample by selecting stars with periods within 5\% of the Ro~=~1.3 curve (which traces the center of the highest density contour in Figure~\ref{fig:kde}), 5500~K~$<$~\teff~$<$~6500~K, and 4.1~dex~$<$~\logg~$<$~4.75~dex. While ages for the \lamostkep sample are not available, the broad distribution of these stars in the spectroscopic H-R diagram supports the inference from the CKS sample that the long-period pileup is populated by stars with a broad range of ages (Figure~\ref{fig:ages}). Interestingly, the solar \teff and \logg values appear to be wholly consistent with the distribution of long-period pileup stars. We discuss the Sun in context of the long-period pileup further in \S\ref{subsec:thesun}.

\begin{figure*}
    \includegraphics[width=\linewidth]{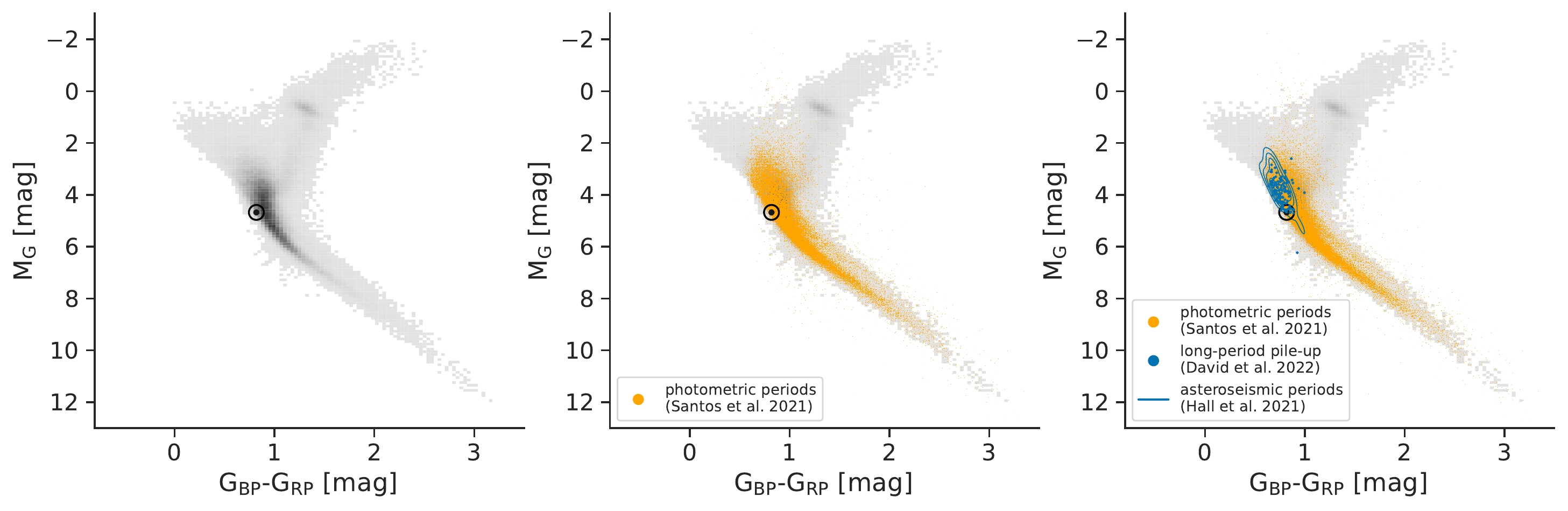}
    \caption{Gaia DR2 color-magnitude diagram. In each panel, a 2-d histogram of all Kepler targets is shown in grayscale, with darker shades representing a higher number of targets in each cell. The estimated position of the Sun is indicated by the black point, based on the calibration of \citet{Casagrande2018}. In the middle and right panels, the \santos targets with photometric rotation period detections are overplotted in orange. In the right panel, the CKS long-period pileup stars identified in this work are shown by blue points and a Gaussian kernel density estimation of the \hall asteroseismic sample is shown by blue contours. No reddening corrections were performed to the photometry in this figure.}
    \label{fig:cmd}
    \script{cmd.py}
\end{figure*}

To further assess the evolutionary state of stars on the long-period pileup, we constructed a color-magnitude diagram (CMD) from the Gaia DR2 photometry and parallaxes for the Kepler field, the \mma and \santos rotation period catalogs, the \hall asteroseismic sample, and the CKS long-period pileup stars from this work (Figure~\ref{fig:cmd}). Stars with detected periods from rotational brightness modulations are primarily solar-type and lower-mass dwarfs, with a comparatively small number of subgiants. The \santos catalog contains more stars, in part due to the higher sensitivity to more slowly rotating, evolved stars, relative to \mma. CKS stars on the long-period pileup clearly occupy a well-defined region of the upper main-sequence which overlaps well with the \hall asteroseismic sample.

\subsection{Where do the pileups end?} \label{subsec:extent}
It appears from Figures~\ref{fig:kde} and \ref{fig:models} that the number density of stars on the long-period pileup declines towards cooler \teff, as predicted by the WMB model (see Figure~13 of \jvs). The number density of stars on the short-period pileup similarly declines towards cooler \teff. If the short-period pileup is due to core–envelope coupling, one might expect an opposite trend of increasing number density toward cooler \teff, since the core–envelope coupling timescales and hence the ``stalled'' braking phases are longer for lower-mass stars \citep{Curtis2020}. However, the observed \teff--\prot distribution depends sensitively on the selection functions and observational biases inherent to both Kepler and the source of \teff (e.g. LAMOST), which likely explains the observed number density trend. 

\begin{figure}
    \centering
    \includegraphics[width=\linewidth]{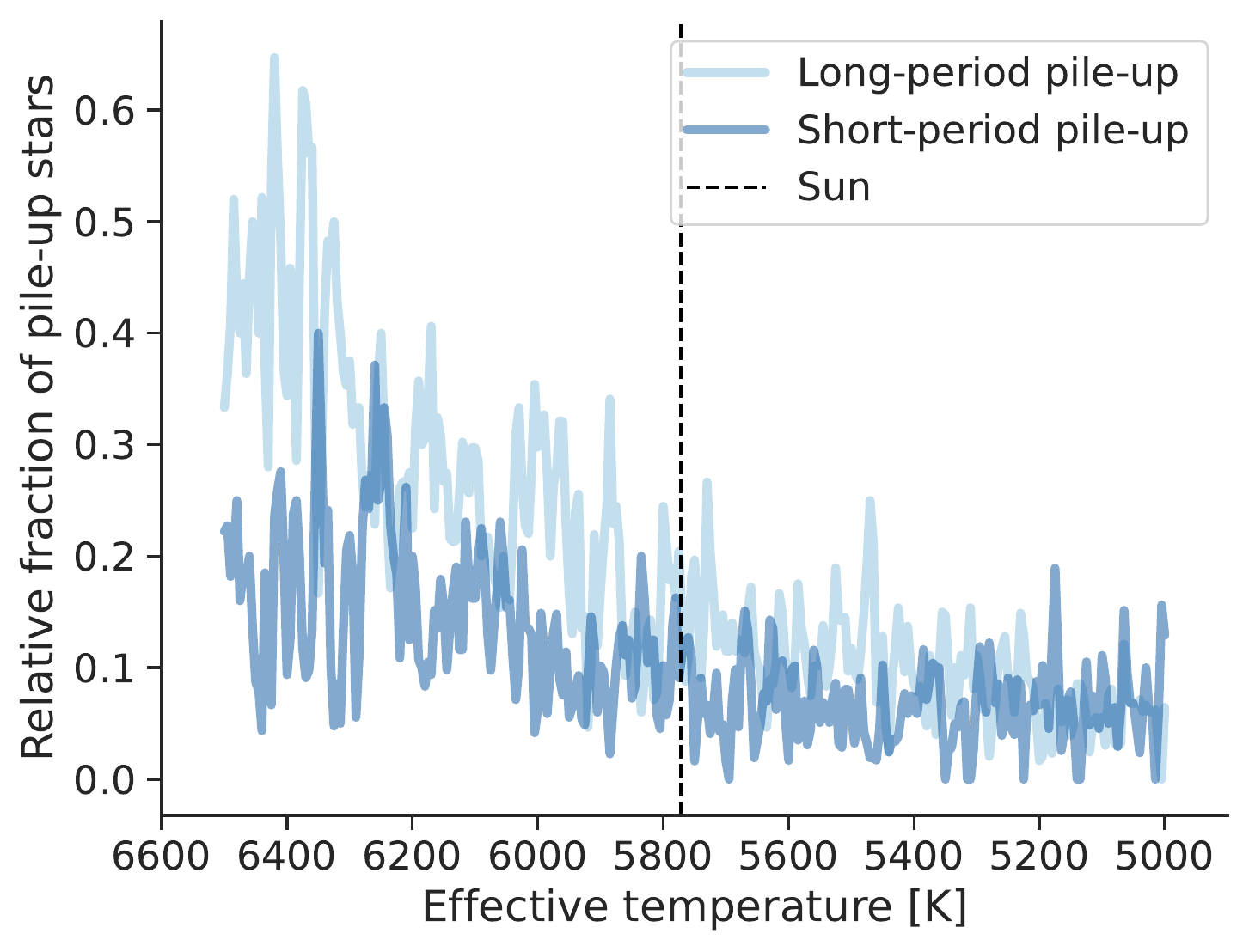}
    \caption{The relative fraction of stars on the long- and short-period pileups, in the \lamostmcq sample, in 10~K windows of \teff (for stars with \logg~$>$4.1 dex).}
    \label{fig:fraction}
    \script{fraction.py}
\end{figure}

While it not clear whether or not these declines are astrophysical in nature, the result of the selection functions or observational biases inherent to Kepler or LAMOST, or some combination of effects, we attempted to characterize the extent of the pileups through the following approach. We found through inspection that constant Rossby curves of Ro=0.5 and Ro=1.3 (for the short- and long-period pileups, respectively) appear to describe the highest density contours found from Gaussian kernel density estimation of the \lamostmcq sample. In windows of 10~K width we measured the fraction of stars with periods within 1~d of each of the two constant Rossby curves, relative to the total number of stars in that \teff window. We found that the relative fraction of stars on the long-period pileup declines rapidly between 6500~K and 5800~K, by more than half in that temperature range before plateauing at cooler \teff (Figure~\ref{fig:fraction}). The relative fraction of stars on the short-period pileup declines more slowly, and below the temperature of the Sun, the relative fractions of stars on the two pileups are nearly equal. Thus, it is not yet clear if either pileup extends to temperatures cooler than \tsun.

\newpage
\section{Discussion} \label{sec:discussion}

\subsection{Detection biases}
\label{subsec:detectionbias}
To this point, we have not addressed the detection biases inherent to Kepler rotation period catalogs. As periods become longer, they become more difficult to measure directly from Kepler time series. This is true for at least three reasons: (1) photometric variation amplitude declines with increasing Rossby number, (2) the existence of significant spot groups appears to be intermittent for high Rossby numbers, and (3) the observational baseline is finite. We note that the last point is not so important for Kepler's long baseline ($\sim$1400~d), while the first two points are due to astrophysical reasons. This is important because, if a period detection threshold is Ro-dependent and the detection threshold Ro$_\mathrm{thresh}$ is close to the value \rocrit at which WMB becomes important, then the location of the pileup in the catalogs considered here depends sensitively on how detectability depends on Ro. For example, if Ro$_\mathrm{thresh}$ is only slightly larger than \rocrit and the long-period pileup has some non-negligible width in period space (due either to astrophysics or measurement error), then it's possible the current data set only reveals the lower edge of the long-period pileup.

\jvs addressed detection biases in detail, finding that a threshold of Ro$_\mathrm{thresh} \approx 2$ could reproduce the observed upper edge of the \mma \teff–\prot distribution. In this scenario, if \rothresh $\lesssim$ \rocrit, it becomes difficult to distinguish between the standard and WMB models in the \mma sample. However, those authors presented arguments that suggest detection bias is not solely responsible for producing the observed \teff–\prot distribution. We direct readers to \S4.2 of that work for a summary of those arguments. Critically, we note here that while an Ro-dependent detection threshold may reproduce the upper edge of the \teff–\prot distribution, it would not be responsible for producing a pileup or overdensity at this edge. To our knowledge, there are no systematics of the rotation period catalogs considered here that would bias stars towards the detection threshold.

\begin{figure}
    \centering
    \includegraphics[width=\linewidth]{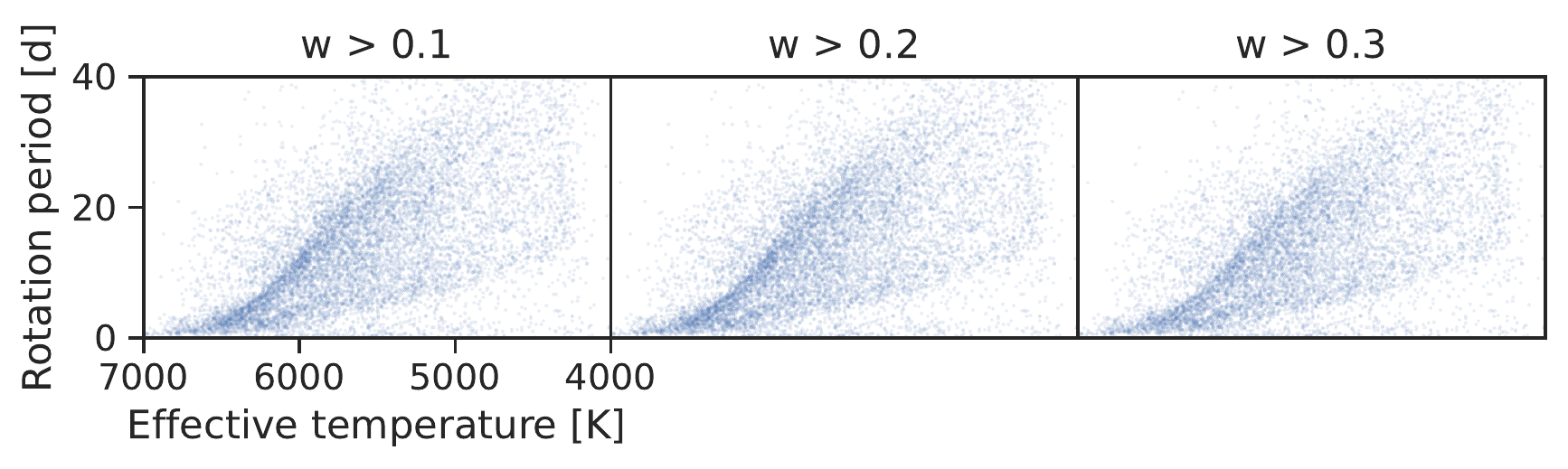}
    \caption{The \teff–\prot distribution of the \lamostmcq sample for different weight thresholds (indicated above each panel) computed by \mma.}
    \label{fig:weights}
    \script{weights.py}    
\end{figure}

To understand whether the period detections on the long-period pileup are reliable, we examined the weight parameters published by \mma. The weight, $w$, is a metric \mma designed to serve as a proxy for the reliability of a rotation period detection, based on a star's autocorrelation function (ACF) properties and its position in $T_\mathrm{eff}$–$P_\mathrm{rot}$–LPH space, where LPH designates the local peak height of the ACF. Those authors selected $w>0.25$ for an acceptable compromise between real detections and false positives. In Figure~\ref{fig:weights}, we show that the long-period pileup is apparent even for $w>0.3$, and we conclude that the majority of stars on the long-period pileup are likely to be genuine detections and not an artifact of some detection threshold.

Most importantly, two independent samples with different observational biases than the surveys considered here also appear to yield a pileup of stars at the long-period edge. These samples are the \hall Kepler asteroseismic sample, which measures rotation through asteroseismic mode splitting (discussed in \S\ref{subsec:asteroseismic}), and the \masuda sample of Kepler stars with Keck/HIRES spectroscopy, from which \masuda inferred the peak of the \teff–\prot distribution using spectroscopically-determined \vsini and constraints on the stellar radii.

We do note, however, that we can not rule out the possibility that the long-period pileup is wider in period or Rossby space than we observe, with stars just above the pileup being undetected through rotational brightness modulations. Such a scenario might result if magnetic braking is gradually weakened once stars reach \rocrit, as opposed to ceasing entirely at \rocrit. However, even in this scenario there are observational constraints on the rate at which the braking index changes. If the braking index starts to decline as Ro approaches \rocrit then there may be tension with the open cluster data. Additionally, if the pileup does have a broader width in Rossby number, the close agreement between our observations and the \hall and \masuda samples suggests the true pileup width cannot be much larger than we observe.

\subsection{Long-period behavior} \label{subsec:longperiod}

The existence of a pileup of stars with a constant Rossby number is consistent with expectations from the WMB model of \citet{vanSaders2016, vanSaders2019}, as discussed in \S\ref{subsec:rossby} and \S\ref{subsec:models}. The location of the long-period pileup coincides closely with the cluster of main-sequence stars with asteroseismic rotation rates determined by \hall, as well as the peak of the period distribution inferred by \masuda from combining \vsini and stellar radii. We suggest that these features are one and the same.

The modest differences between the long-period pileup we observe and the features observed by \hall and \masuda might be explained by differing \teff scales and differing observational biases of the samples. The present samples have rotation periods detected from photometric variations and, for a survey of finite baseline and sensitivity like Kepler, photometric rotation periods are harder to detect for slower rotators, smaller amplitude variations, and more stochastic variability patterns (see \S\ref{subsec:detectionbias}). The \hall asteroseismic sample and \masuda sample, by comparison, are not biased against quiet, unspotted stars and are more likely to contain pileup stars that our sample may have missed.

In some cases, the \rocrit values we inferred in \S\ref{subsec:rossby} were smaller than the value of \rocrit~$\approx$~\rosun~$\approx$~2 found by \citet{vanSaders2016,vanSaders2019}, sometimes with high statistical significance. We caution against overinterpreting the \rocrit values inferred here, and enumerate below the ways in which the \rocrit inference may be biased.

\begin{enumerate}
    \item The \taucz relation adopted impacts the inferred \rocrit. We explored the \taucz formulae of \citet{BarnesKim2010}, \citet{Landin2010}, and \citet{Amard2019}, but these produce larger \taucz at fixed \teff than the \citet{CranmerSaar2011} relation, leading to even smaller inferred \rocrit values. We also explored using a fit to the \taucz computed in \jvs for solar-metallicity stars, which produces a smaller \taucz at fixed \teff than the \citet{CranmerSaar2011} relation. This relation revises our \rocrit estimates upwards by 15\%. 
    \item While our \rocrit inference procedure includes a constant \teff offset between the data and the \taucz relation, it does not include any \teff-dependent offsets. In Appendix~\ref{app:teff}, we show that the LAMOST \teff appear to have a non-linear mapping to the other \teff scales considered here. Perhaps not coincidentally, the \lamostmcq sample favors a \rocrit that is the least compatible with other samples. We also note that the MCMC results for the \lamostmcq sample preferred the LAMOST \teff to be shifted lower (Table~\ref{tab:mcmc}), which is in inconsistent with our finding that the LAMOST \teff are already lower than those from other spectroscopic surveys (Appendix~\ref{app:teff}). 
    \item Defining the long-period pileup or edge is a choice that impacts the inferred value of \rocrit. For example, though we chose the 90th \prot percentile in the \lamostmcq sample, \jvs chose the 95th percentile. Similarly, our boundary for the CKS long-period pileup was chosen subjectively.
\end{enumerate}

It is evident from Figure~\ref{fig:gap} that the long-period pileup also coincides with a steep gradient in variability amplitude, such that a hidden population of pileup stars may lie just beyond the edge of detectability. This effect, in addition to the effects mentioned above, can further bias our inference of \rocrit to lower values. Regardless of the biases mentioned above and the true value of \rocrit, the sample considered here clearly indicates that a pileup exists at the long-period edge for stars with \teff~$\gtrsim$~\teffmin. By construction, this pileup occurs at a Rossby number lower than that associated with a detection edge. In other words, the fact that we observe a pileup indicates that if WMB is assumed to be the cause, then \rocrit~$\lesssim$~\rothresh, where \rothresh again is assumed to be a detection threshold.

We also note that the WMB model makes the simplistic assumption that magnetic braking ceases when Ro reaches \rocrit. However, it is possible that magnetic braking becomes gradually weaker, a scenario which would also lead to a pileup but may also cause such a pileup to have some intrinsic width which is not due to measurement uncertainties.

\begin{figure}
    \centering
    \includegraphics[width=\linewidth]{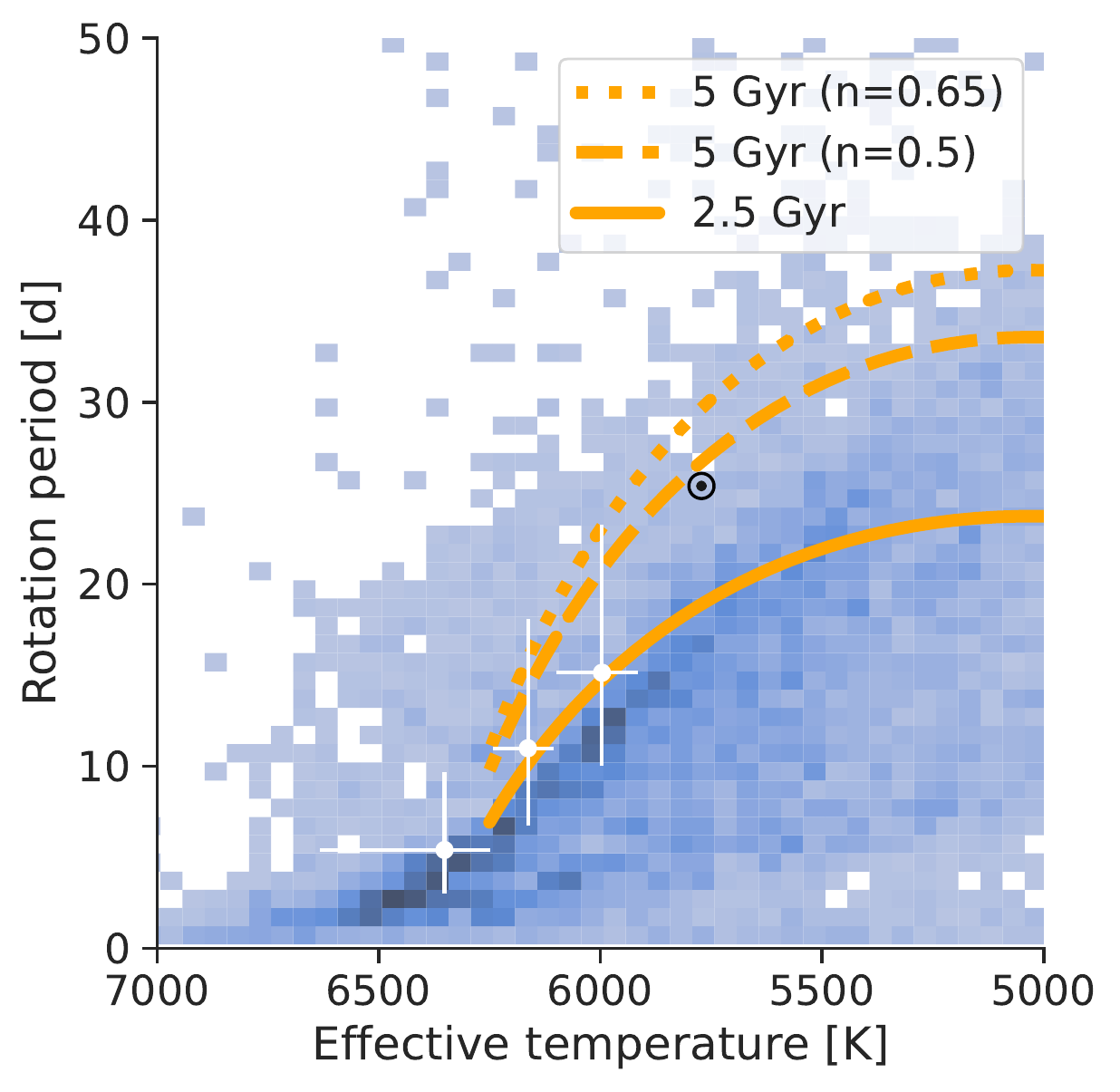}
    \caption{A 2-d histogram of the \lamostmcq \teff–\prot distribution compared to an empirical $\approx$~2.5~Gyr hybrid cluster sequence from \curtis (solid line). The dashed and dotted lines show the same sequence evolved forward to 5~Gyr assuming braking indices of $n=0.5$ (Skumanich-like spin-down) and $n=0.65$, which is favored by \curtis. For comparison, the white points indicate the peak of the \prot distribution from \masuda and the black point indicates the Sun.}
    \label{fig:skumanich}
    \script{skumanich.py}    
\end{figure}

Our observation of a long-period pileup stands in contrast with recent rotation studies of solar analogs. \citet{doNascimento2020} studied the rotation period distribution of 193 solar analogs, concluding that some solar-mass main-sequence stars appear to rotate significantly more slowly than the Sun, seemingly at odds with the WMB model. Additionally, \citet{LorenzoOliveira2019} studied the \prot--age relation of solar twins observed by the Kepler mission, finding marginal statistical evidence in favor of a standard spin-down model over the WMB model. Those authors posited that if WMB takes place for Sun-like stars, it should happen at \rocrit~$\gtrsim 2.29$ or ages~$\gtrsim$~5.3~Gyr. A case study of an $\sim$8~Gyr solar twin further reinforced these conclusions \citep{LorenzoOliveira2020}.  By comparison, our findings provide support for the WMB model among stars that are slightly hotter than the Sun, but at \rocrit~$\lesssim$~\rosun and at ages in the range of $\sim$2--6~Gyr. In contrast to the findings of those authors, we find that the long-period pileup lies far below an empirical 2.5~Gyr gyrochrone of \curtis that is evolved forward to 5~Gyr for braking indices of $n=0.5$ or $n=0.65$ (Figure~\ref{fig:skumanich}). This implies the braking index must drop to a much lower value at some time after 2.5~Gyr for Sun-like stars.

\hfill\break
\subsection{Short-period behavior} \label{subsec:shortperiod}

Unlike the long-period pileup, the short-period pileup is not predicted by the WMB model or, more generally, any standard, solid-body braking model. However, this feature may also be due to an epoch of apparent stalling, albeit a temporary one since cluster studies demonstrate that stars continue to spin down beyond the short-period pileup. Works examining the \teff–\prot sequences of open clusters have found overlap between low-mass members in clusters of different ages \citep{Agueros2018, Curtis2019a, Curtis2020}, notably Praesepe (0.67~Gyr), NGC 6811 (1.4~Gyr), and NGC 752 (1.4~Gyr). In other words, the spin rates of low-mass stars appear to evolve very little in the time elapsed between the ages of those clusters. The short-period pileup we observe may be the manifestation of the same type of slowed spin evolution, but for stars of higher masses and younger ages than the cluster members in the above-mentioned works (since the pileup is observed at shorter periods relative to the Praesepe sequence). 

\citet{Curtis2019a, Curtis2020} proposed that the overlapping cluster sequences could be produced by a temporary epoch of stalled spin-down, caused either by (i) a reduction in the magnetic braking torque, or (ii) core–envelope momentum transfer which offsets the effect of magnetic braking \citep[e.g.][]{MacGregor1991}. In the core–envelope momentum transfer scenario angular momentum is exchanged between the envelope and the core on a characteristic timescale known as the core–envelope coupling timescale. The angular momentum transfer spins up the envelope, temporarily offsetting the spin-down via magnetic braking. Thus, in the core–envelope coupling scenario, spin evolution is slowed when the star's age becomes comparable to the core–envelope coupling timescale. Theoretical predictions for the core–envelope coupling timescale for a solar-mass star range from 30--110 Myr \citep{MacGregor1991,Krishnamurthi1997, Bouvier2008, IrwinBouvier2009, Denissenkov2010, GalletBouvier2015, Lanzafame2015, Somers2016, Spada2020}. 

After a period of slowed spin-evolution, stars must resume spinning down as evidenced from studies of older open clusters. From open cluster data \curtis estimated that solar-mass stars resume spin-down after an age of $\approx$~230~Myr. Thus, if core–envelope coupling is responsible for delaying the spin-down of stars, and if the theoretical core–envelope coupling timescales are accurate, then we may expect Sun-like stars to experience slowed spin evolution between $\sim$100~Myr and $\sim$200~Myr. 

Both the theoretically predicted core–envelope coupling timescale and the observationally inferred timescale for the resumption of spin-down are consistent with our observation that the short-period pileup is intermediate to the Pleiades (0.12~Gyr) and Praesepe (0.67~Gyr) cluster sequences. Furthermore, in order to match observations of rotation periods in young clusters, models require that the core–envelope coupling timescale increases towards lower stellar masses \citep[e.g.][]{Irwin2007, Denissenkov2010, GalletBouvier2015}, which provides a natural explanation for why this stalling occurs at older ages for lower-mass stars in \curtis. 

We emphasize that the temporary epoch of slowed spin-down (also referred to as stalled magnetic braking) proposed by \citet{Curtis2019a, Curtis2020} is not to be confused with the termination of magnetic braking that characterizes the WMB model of \citet{vanSaders2016, vanSaders2019}. The physical mechanisms thought to be responsible for each of these proposed stages of rotational evolution are distinct, though it is interesting that both the long- and short-period pileups seem to be well-described by curves of constant Rossby number. As mentioned above, one theory for the earlier stage of stalled spin-down is core–envelope coupling. Crucially, in the stalled spin-down phase, wind-driven angular momentum losses are not ceased but rather offset by the spin-up torque from core–envelope coupling. In contrast, in the WMB scenario, wind-driven angular momentum losses are proposed to cease ($dJ/dt$=0), with subsequent rotational evolution dictated by the changes in the moment of inertia. 

Finally, we note that we regard the short-period pileup as less secure than the long-period pileup. Preliminary tests with other density estimation methods \citep{Contardo2022}\footnote{\url{https://github.com/contardog/FindTheGap}} also reveal the short-period pileup, but further work is needed to better understand this feature. In Appendix~\ref{app:harmonics} we show that the short-period pileup does not appear to be a harmonic of the long-period pileup, as one might expect if the periods on the short-period pileup were erroneously determined. However, the short-period pileup is primarily observed in the \lamostmcq sample, while it is unclear whether it is present in the \lamostsan sample (see the rightmost panels of Figure~\ref{fig:xmatch}). The differences between the \mma and \santos catalogs may provide an explanation for this observation. The latter catalog contains a much larger number of detections at longer rotation periods, and the higher sensitivity to more slowly rotating stars within the \santos may then make the short-period pileup appear weaker relative to the long-period pileup. This is as expected since open cluster observations suggest that the short-period pileup can not be a long-lived feature, while the long-period pileup appears to be much longer lived. Consequently, the \santos catalog may more accurately reflect the relative strengths of these two features.

\subsection{Implications for the period gap} \label{subsec:gap}

An unexplained feature of the Kepler rotation period distribution is the existence of a bimodal period distribution for dwarf stars of similar \teff. The effect was first noticed for M-dwarfs \citep{McQuillan2013b}, but was later shown to extend to $\sim$5000~K \citep{Reinhold2013, McQuillan2014, ReinholdHekker2020}, and even to $\sim$6500~K \citep{Davenport2017}. 

\citet{McQuillan2013b, McQuillan2014} speculated that this period bimodality could originate from stellar populations of different ages, an explanation seemingly supported by a correlation between the strength of the bimodality and height above the galactic disk \citep{DavenportCovey2018}. However, \curtis demonstrated that cluster sequences cross the gap, invalidating the claim that the feature is caused at a specific age, as one might expect from a period of decreased star formation. Additionally, \citet{Gordon2021} found that the gap is observed across the many fields observed by the K2 mission, which is in tension with the star formation history hypothesis as different Galactic sight lines are expected to have different different star formation histories (for sufficiently large volumes). 

An alternative explanation for the gap was proposed by \citet{Reinhold2019}, who found that the dearth of stars with intermediate rotation periods is associated with a decrease in photometric variability. Consequently, those authors proposed that the period bimodality may be the result of a transition between spot- and faculae-dominated photospheres. In this scenario, the period gap is due to bright faculae canceling out the effects of dark star spots. 

\begin{figure*}
    \centering
    \includegraphics[width=\linewidth]{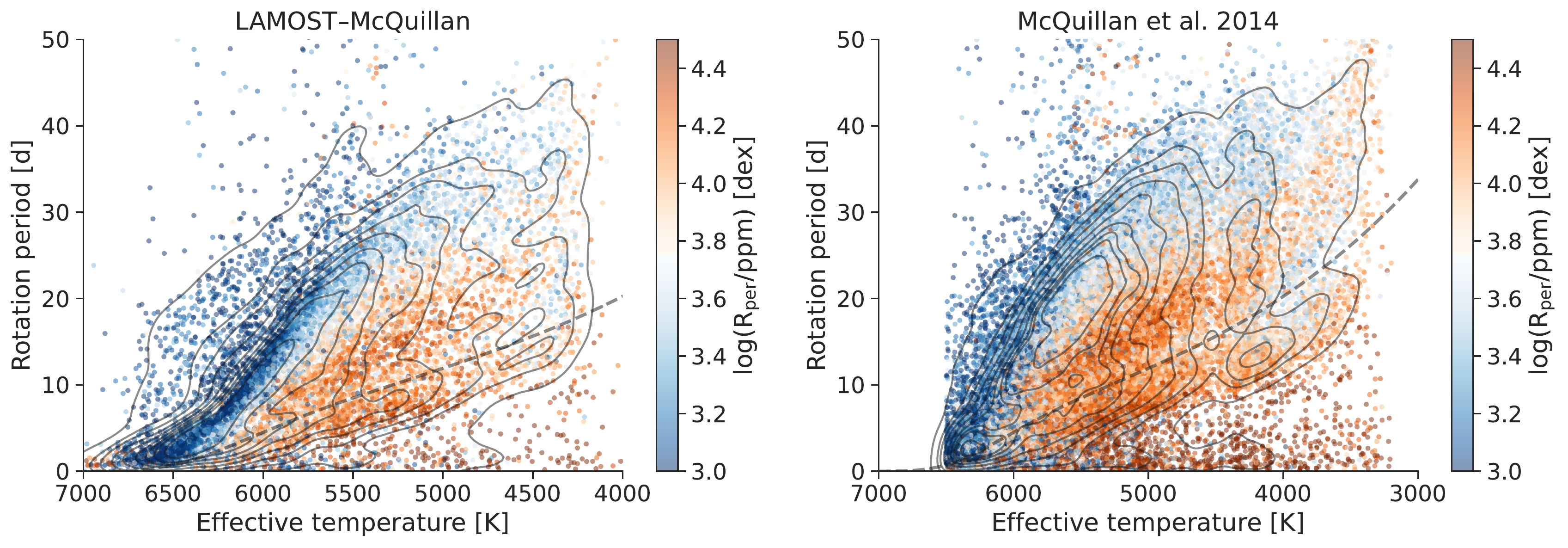}
    \caption{The \teff--\prot distribution of the \lamostmcq (left) and \mma samples color coded by the variability amplitude, \rper. Black contours show Gaussian kernel density estimation of the plotted distributions, and the dashed line shows a constant Rossby curve of Ro~=~0.5.
    The long- and short-period pileups are separated by a relative dearth of stars with intermediate rotation periods. A strong gradient in \rper is apparent across this gap, such that variability amplitude reaches a local minimum near the gap's center.}
    \label{fig:gap}
    \script{gap.py}
\end{figure*}

The short- and long-period pileups we examine here naturally produce a dearth of rotators at intermediate periods. This gap is the same period gap noticed by the authors mentioned above, as made apparent when comparing the \lamostmcq sample to the original \mma sample. Moreover, as seen in Figure~\ref{fig:gap}, we recover the gradient in photometric variability across the gap pointed out by \citet{Reinhold2019}. The photometric variability, as measured through the \rper metric published by \mma, reaches a local minimum near the location of the gap. This supports the notion that the cause of the gap is due to changes in the stars themselves, rather than being the result of mixed stellar populations. However, the variability levels on both sides of the gap are not close to the detection limit, as evidenced by the fact that periods are securely detected for stars with similar properties at much lower \rper values. This would suggest that the period gap is not due solely to a detection issue, unless variability levels were to drop precipitously as stars approached the gap. 

\citet{Gordon2021} proposed that the gap could instead be due to a period of accelerated spin-down immediately proceeding the stalling due to core–envelope coupling (such that stars evolve quickly through the gap and are rarely observed there). We can neither confirm nor reject this scenario, and we note that the gap is significantly emptier when using Gaia colors \citep{DavenportCovey2018, Gordon2021} compared to when using spectroscopic \teff as we do here (see also Appendix~\ref{app:gaia}). Notably, if the short-period pileup is indeed due to core–envelope coupling, and the gap due to a period of accelerated spin-down after such a coupling episode, the observed gradient in photometric variability still requires a physical explanation.

\hfill\break
\subsection{Does the Sun reside on the long-period pileup?} \label{subsec:thesun}

It is unclear whether or not the Sun is a resident of the long-period pileup. From Fig.~\ref{fig:xmatch} it is clear that the long-period pileup extends to temperatures as cool as the Sun's. If one assumes there are no systematic offsets between the spectroscopic \teff scales considered here and that of the Sun, the Sun's equatorial rotation period places it $\sim$5--7 days above the long-period pileup. This raises the question of how the Sun's angular momentum has evolved to its current state, and whether WMB is a generic evolutionary phase. 

If the Sun is indeed on the long-period pileup, it likely the presence of \teff systematics and/or observational biases that obfuscate that fact. To place the Sun on the long-period pileup would require shifting the spectroscopic temperatures higher by 111~K for the CKS sample, 143~K for the SPOCS sample, or 116~K for the LAMOST sample. We also note that there are \teff-dependent systematics in the LAMOST \teff that, if corrected for, would bring the \lamostmcq long-period pileup into closer agreement with a solar Rossby curve. Comparing the Sun to the \hall asteroseismic sample (Figure~\ref{fig:kde}) seems to support the notion that the Sun may indeed reside on the long-period pileup, or very close to it, without the need for a large \teff shift. This apears to be at odds with the notion that a \teff shift alone can place the Sun on the long-period pileup, since the \hall and CKS temperature scales agree to within $\sim$30~K. However, it is still possible that there are \teff-dependend systematics in the temperature scales.

Furthermore, as discussed in \S\ref{subsec:longperiod}, the \hall sample is less prone to detection bias which acts to censor stars with higher Ro and lower variability amplitudes in the samples considered here. As shown by \citet{Aigrain2015}, the pipelines used to measure rotation periods from Kepler data would not be guaranteed to detect the solar rotation period. Regardless, it is clear from the comparisons in \S\ref{subsec:asteroseismic} that the long-period pileup can not be much broader (in \prot or Rossby space) than we observe.

\section{Conclusions} \label{sec:conclusions}

Our primary conclusions are summarized as follows:

\begin{enumerate}
    \item We observe an overdensity at the long-period edge of the \teff-\prot distribution of Kepler main-sequence stars with \teff~$\gtrsim$~\teffmin. We hypothesize that this pileup was previously obfuscated by imprecise \teff estimates. Both the existence of the pileup and its obscuration by large \teff errors were predicted by \citet{vanSaders2019} as a consequence of weakened magnetic braking for stars with $M\gtrsim$1~\msun. 
    
    \item The long-period pileup is well-described by a constant Rossby number, with a critical value of \rocritfinal, in the \teff range of $\approx$~5500-6250~K. The pileup is also populated by stars with a wide range of isochrone ages (\agerange). A pileup of stars with a constant Rossby number and a broad range of ages is a prediction of the WMB model of \citet{vanSaders2016, vanSaders2019}. The precise value of \rocrit is sensitive to \teff scale shifts between observational data and the models used to compute \taucz. 

    \item Comparison of the long-period pileup with empirical rotation sequences from open clusters implies that stars with $M\gtrsim1$~\msun pile up onto the ridge on a timescale $>$~1~Gyr but $\lesssim$2.5~Gyr, compatible with the predictions of \jvs. Using isochrone ages for a sample of exoplanet hosts on the long-period pileup suggests that stars slightly hotter than the Sun may populate the pileup until an age of $\sim$6~Gyr.
    
    \item It is yet unclear whether the Sun resides on the long-period pileup or has already evolved through it. Offsets of $\approx$110--140~K between the observational and theoretical \teff scales would place the long-period pileup at Rossby numbers consistent with other literature values of the location of the WMB transition. If the Sun has evolved through the pileup, there is some modest tension with the Sun's age and the ages of the oldest stars on the pileup; some of the cooler long-period pileup stars in the CKS sample are both more massive and older than the Sun, which contradicts the expectation from the WMB model that hotter stars spend a shorter period of time on the pileup. However, this tension might be simply explained by inaccurate isochrone ages.
    
    \item We tentatively detect a secondary overdensity of stars at the short-period edge of the \teff-\prot plane. This overdensity appears to be less prominent than the long-period overdensity, possibly indicating that the short-period pileup is shorter-lived, $\mathcal{O}(10^8 \text{yr})$, relative to the long-period pileup ($>10^9$~yr). The short-period pileup appears to be intermediate to the empirical Pleiades (0.12~Gyr) and Praesepe (0.67~Gyr) open cluster sequences and may result from a temporary epoch of stalled spin-down due to core–envelope coupling, as proposed by \curtis. The short-period pileup can also be fit with a constant Rossby model, though over a range of \teff that differs from that of the long-period pileup.
    
    \item The number density of stars on the long-period pileup declines with \teff, in line with predictions from the WMB hypothesis, though it remains unclear whether this observation is due to astrophysics, the Kepler selection function, observational biases, or some combination of effects. The relative fraction of stars on the long-period pile up declines by a factor of $\sim$2 between $\sim$6200~K and $\sim$5800~K. 
    
    \item We find tentative evidence for an age-gradient along the long-period pileup, such that hotter stars on the ridge are younger on average. Relatedly, the age dispersion along the ridge is non-uniform as a function of temperature, with hotter stars showing a smaller dispersion. This observation suggests that hotter stars reside on the long-period pileup for a shorter period of time relative to cooler stars. These observations are in accordance with predictions from the WMB model, which predicts that stars of different masses spend an approximately equal fraction of their respective main-sequence lifetimes on the long-period pileup. However, a more careful analysis is required to conclusively show these observations are not due to the intrinsic age gradient expected among a sample of main sequence stars with different masses and the higher isochrone age uncertainties associated with cooler stars.
    
    \item The existence of the long-period pileup limits the utility of gyrochronology for the hottest stars with convective envelopes, as stellar spin-down appears to stall on the pileup. For example, a Sun-like star may spend several Gyr evolving through the long-period pileup. Authors using gyrochronology as a means of age-dating a field dwarf star with \teff~$\gtrsim$~\teffmin (and possibly cooler \teff as well) should take care to assess whether that star resides on the long-period edge, in which case the uncertainty on the age may be larger than current gyrochronology calibrations imply.
    
    \item An increasing number of open clusters are being discovered by searching for a clustering of rotation periods along a slow-rotator sequence in the \teff–\prot or color–\prot plane. However, the long-period pile up discovered here can mimic a slow-rotator sequence in a small sample of unassociated stars with different ages and precisely measured temperatures and rotation periods. This is clearly demonstrated by the CKS sample in the right-hand panel of Figure~\ref{fig:surveys}. Taken out of context, this sample resembles a group of coeval stars with a slow-rotator sequence. The discovery of pileups in stellar rotation periods therefore has consequences for open cluster studies. When an overdensity or ridge is present in the rotation period distribution of a stellar population, care must be taken to ensure that it is not caused by WMB or core–envelope coupling before assuming that population is coeval and using the overdensity to age-date it.
    
\end{enumerate}

The code and data tables required to reproduce the figures and analysis presented here are publicly available through GitHub.\footnote{\url{https://github.com/trevordavid/rossby-ridge}} The data tables are also available through Zenodo.\footnote{\url{https://doi.org/10.5281/zenodo.6391526}}

\appendix
\section{Comparison of temperature-period distributions}\label{app:teffprot}
\begin{figure*}
    \centering
    \includegraphics[width=\linewidth]{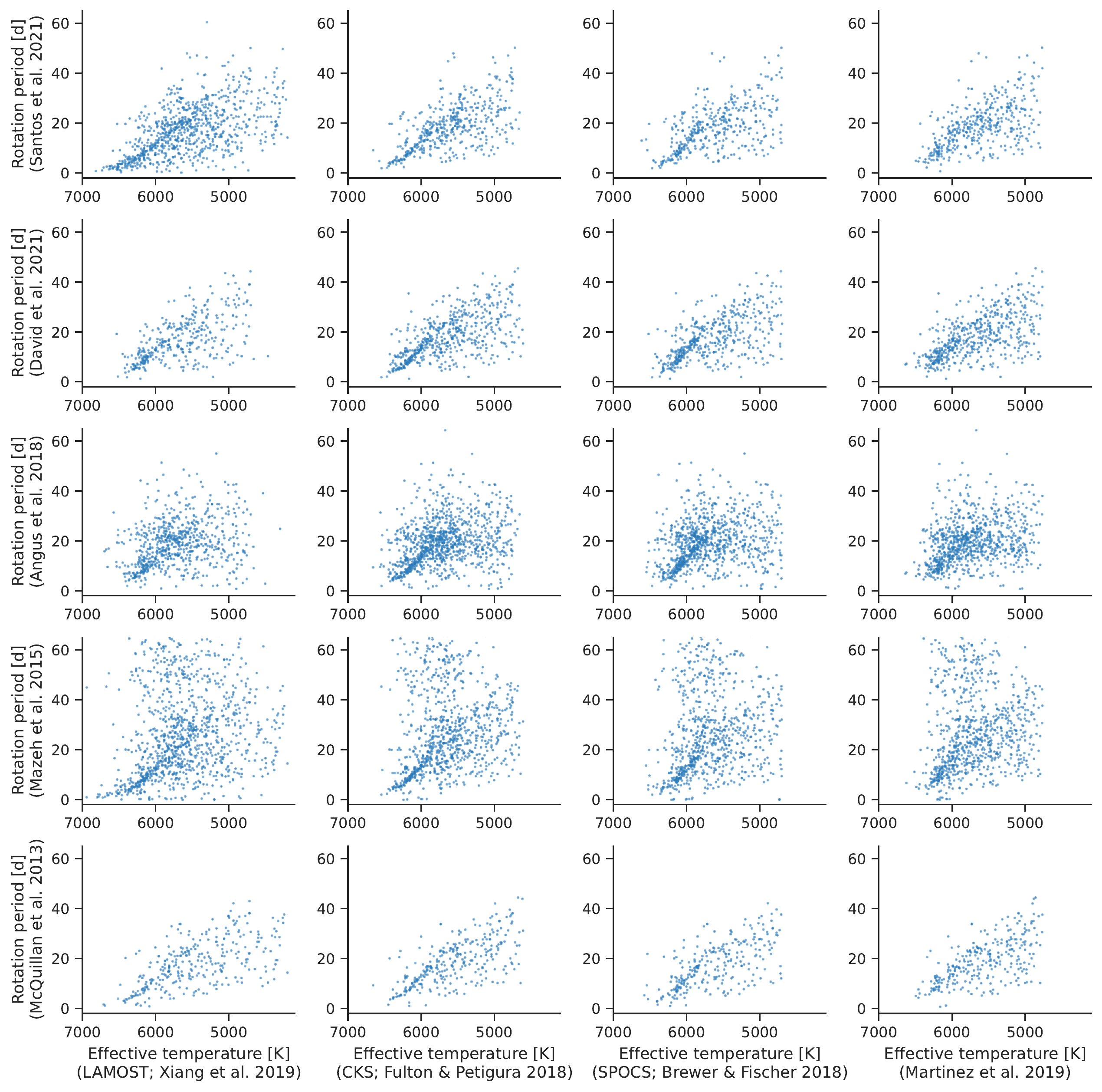}
    \caption{Comparison of the \teff--\prot distribution for the CKS sample using rotation periods and \teff from the sources indicated by the axes labels.}
    \label{fig:comparison}
    \script{comparison.py}
\end{figure*}

In Figure~\ref{fig:comparison}, we show how the \teff--\prot distribution of the CKS sample changes when sourcing \teff and \prot from different, homogeneous catalogs in the literature. The sharpness of the long-period pileup appears to be determined primarily by the source of \teff, rather than \prot. The CKS-Gaia catalog \citep{Fulton2018} appears to offer the highest internal precision.  

\section{Comparison of spectroscopic temperature scales}\label{app:teff}

\begin{figure}
    \centering
    \includegraphics[width=\linewidth]{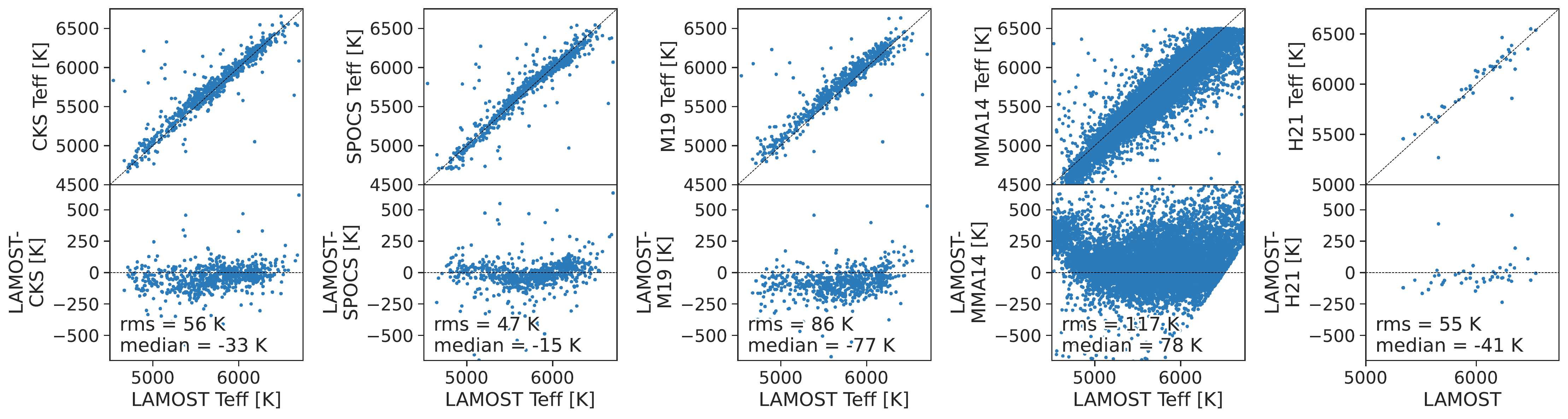}
    \caption{Comparison of \teff estimates from different catalogs: LAMOST \citep{Xiang2019}, CKS \citep{Fulton2018}, SPOCS \citep{Brewer2018}, M19 \citep{Martinez2019}, \mma, and \hall.}
    \label{fig:teffscales}
    \script{teffscales.py}
\end{figure}

In Figure~\ref{fig:teffscales} we compare temperatures between the LAMOST DR5 catalog \citep{Xiang2019} and temperatures from other surveys. We find that the LAMOST \teff scale is consistently cooler than other surveys by $\sim$20--80~K, with the exception of the \mma study which sourced photometric \teff estimates from the KIC \citep{Brown2011}. A LAMOST \teff scale which is systematically cooler provides support to the notion that \rocrit determined from LAMOST temperatures will be systematically underestimated. 

\section{The Gaia color--period plane}\label{app:gaia}
Given that the long-period pileup was previously obscured by imprecise \teff measurements, we explored whether the feature could be recovered using the high-precision photometric colors provided by the Gaia mission \citep{Gaia2016}. We cross-matched the \mma and \santos samples with Gaia EDR3 \citep{GaiaEDR3} using a 1 arcsec search radius and the \texttt{astroquery} package \citep{astroquery}. We retrieved reddening estimates for each target by querying the \texttt{Bayestar19} 3D dust map using the \texttt{dustmaps} package \citep{dustmaps, Green2019}. We then compared the ($G_{BP}-G_{RP}$)–\prot distribution with constant Rossby curves. For this exercise, we used the empirically calibrated \taucz relation of \citet{Corsaro2021}, who presented \taucz as a quadratic function of Gaia $G_{BP}-G_{RP}$ color using the Kepler LEGACY asteroseismic sample as calibrators. We found that the long-period pileup is clearly visible in the ($G_{BP}-G_{RP}$)–\prot plane once stars with high reddening ($A_V>0.2$) are excluded. In order to match a curve of constant Rossby number, given by $\mathrm{Ro}=\mathrm{Ro_\odot}=0.496$ on the \citet{Corsaro2021} scale, we found that a $\approx$-0.1 mag shift to the Gaia colors of the data was required (or, equivalently, a +0.1 mag shift applied to the constant Rossby curve). While there is no justification for such a large shift, it may indicate the presence of a systematic offset in the \taucz relation. We also note that we do not have an explanation for why the solar value Rossby curve does not pass through the Sun.

We note that the morphology of the color–period distribution appears to be different between the Gaia–McQuillan and Gaia–Santos catalogs, with the Gaia–Santos distribution presenting a break near $G_{BP}-G_{RP}= 0.7$. This break is not apparent in the Gaia–McQuillan sample. Similar behavior is seen in \teff–\prot plane, as shown in Figure~\ref{fig:xmatch}, suggesting the origin of the discontinuity is in the rotation periods rather than the temperatures or colors. At present, we do not have a satisfactory explanation for this behavior, though we note that the \santos catalog employed various time series filters which might introduce systematic artifacts.

\begin{figure}
    \includegraphics[width=\linewidth]{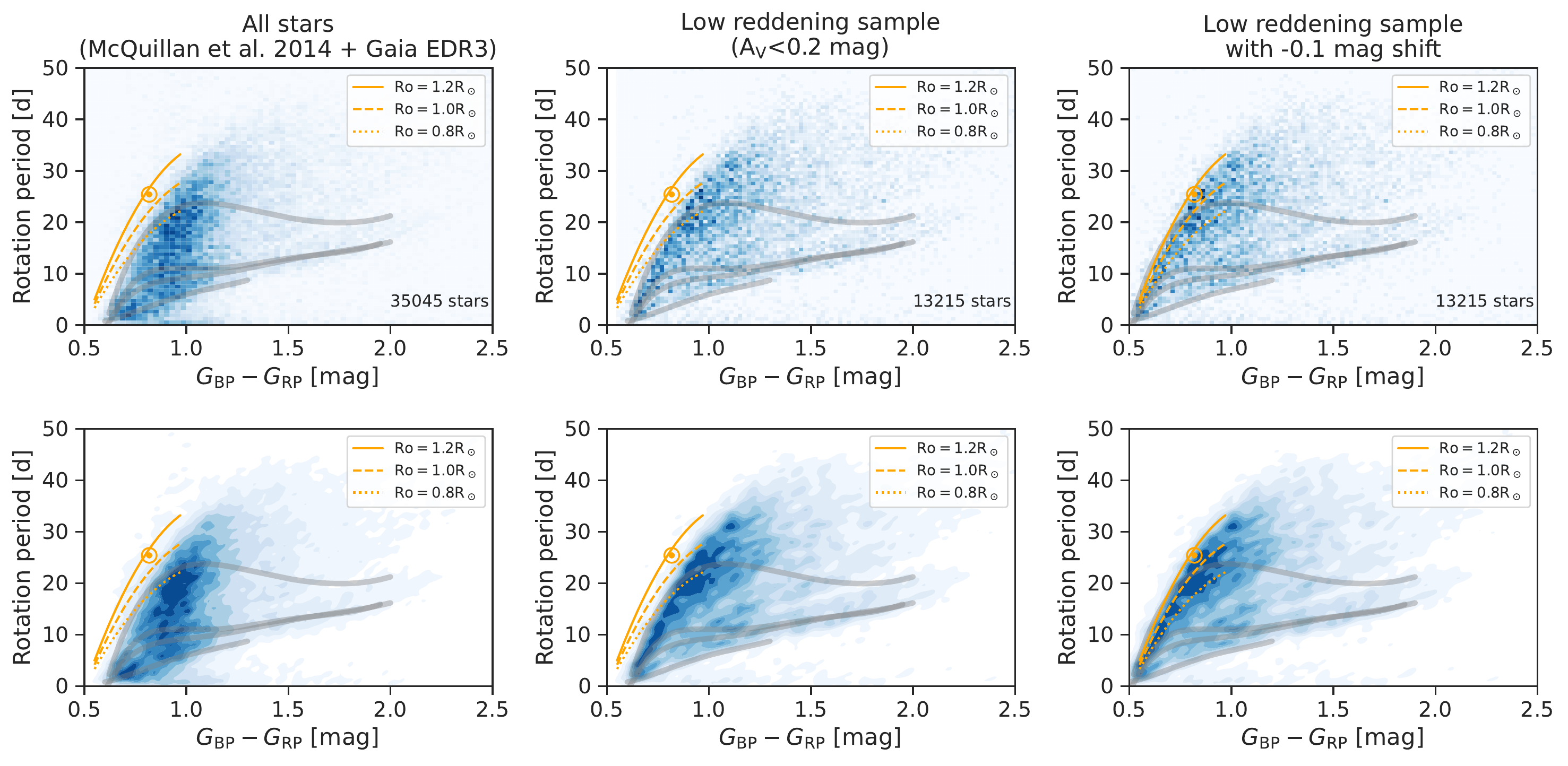}
    \caption{The color–\prot plane for the \mma sample in relation to a curve of constant Rossby number (orange dashed curve, $\mathrm{Ro} = \mathrm{Ro}_\odot = 0.496$) from the empirical \taucz calibration of \citet{Corsaro2021}, which is valid in the color range $0.55 < (G_\mathrm{BP}−G_\mathrm{RP}) < 0.97$. The top rows show 2–d histograms of the probability density for the entire sample (left), a subsample with low reddening (middle), and the same low reddening subsample with a -0.1 mag color shift applied to the data and cluster sequences. Each panel in the bottom row shows a Gaussian kernel density estimation of the respective panel above. The orange point in each panel represents the position of the Sun, using the estimated Gaia color of \citet{Casagrande2018}. The grey lines indicate empirical cluster sequences derived in \curtis. From top to bottom, the cluster sequences are Pleiades (0.12~Gyr), Praesepe (0.67~Gyr), NGC 6811 (1~Gyr), and NGC 6819 + Ruprecht 147 (2.5~Gyr).}
    \label{fig:gaia-mcquillan}
    \script{gaia-mcquillan.py}
\end{figure}

\begin{figure}
    \includegraphics[width=\linewidth]{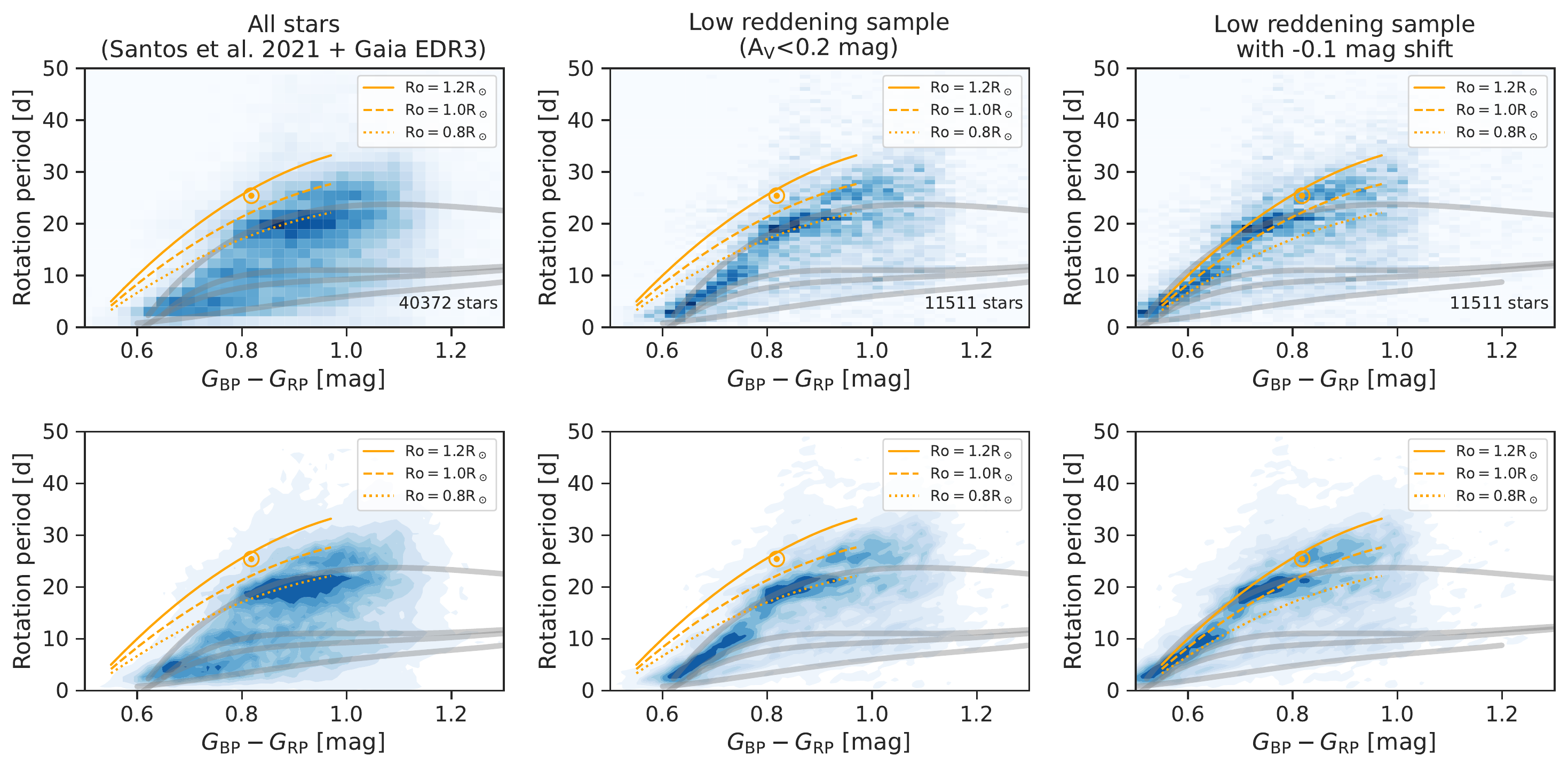}
    \caption{The same as Figure~\ref{fig:gaia-mcquillan} but for the \santos sample.}
    \label{fig:gaia-santos}
    \script{gaia-santos.py}
\end{figure}

\section{Confounding scenarios for the short-period pileup}\label{app:harmonics}

\begin{figure}
    \centering
    \includegraphics[width=0.5\linewidth]{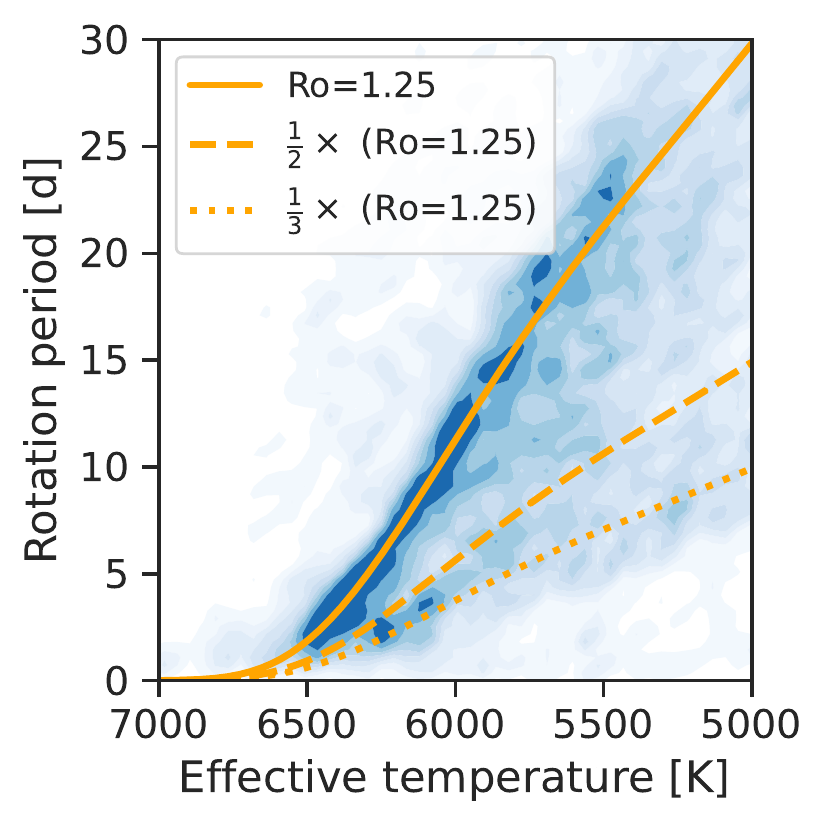}
    \caption{Gaussian kernel density estimation of the \teff–\prot distribution in the \lamostmcq sample for stars with \logg~$>$~4.1. A curve of constant Rossby number is shown by the solid line. The \prot/2 and \prot/3 harmonics of this curve are shown by the dashed and dotted lines, respectively. The short-period pileup is seen near the short-period edge, particularly for \teff~$>$~6000~K. The short-period pileup does not appear to be simply a harmonic of the long-period pileup, as one might expect if the periods for the stars in the short-period pileup were erroneously determined.}
    \label{fig:harmonic}
    \script{harmonic.py}    
\end{figure}

Here we consider the possibility that there is only one astrophysical overdensity in the true \teff–\prot distribution, the long-period pileup, and other features in the observed \teff–\prot distribution appear at period harmonics of this feature due to the difficult problem of reliable, automated rotation period measurement for large samples of stars. Figure~\ref{fig:harmonic} shows a Gaussian kernel density estimation of the \lamostmcq \teff–\prot distribution for stars with \logg~$>$~4.1. The short-period pileup is observed in this sample, particularly for \teff~$>$~6000~K. We found through inspection a constant Rossby curve that traces the long-period pileup. Taking this curve and dividing the periods by factors of 2 and 3, we find that neither resulting curve satisfactorily matches the short-period pileup, though they both bracket the feature seen through density estimation. We interpret this as evidence that the short-period pileup is not simply due to measurement error, although we have not definitively ruled out that scenario.

\begin{acknowledgments}
We thank the anonymous referee for a thoughtful review, as well as Travis Metcalfe and Matteo Cantiello for helpful discussions. We also thank Rodrigo Luger for helpful discussions and assistance with the \texttt{showyourwork!} package, Adrian Price-Whelan for providing the APOGEE--Kepler cross-match catalog, and Karl Jaehnig for plotting advice. It is a pleasure to thank the Stars Group at the American Museum of Natural History, and the Astronomical Data Group at the Flatiron Institute for helpful discussions. 

This work made use of the gaia-kepler.fun crossmatch database created by Megan Bedell. This paper includes data collected by the Kepler mission and obtained from the MAST data archive at the Space Telescope Science Institute (STScI). 

Funding for the Kepler mission is provided by the NASA Science Mission Directorate. STScI is operated by the Association of Universities for Research in Astronomy, Inc., under NASA contract NAS 5–26555. 

Guoshoujing Telescope (the Large Sky Area Multi-Object Fiber Spectroscopic Telescope LAMOST) is a National Major Scientific Project built by the Chinese Academy of Sciences. Funding for the project has been provided by the National Development and Reform Commission. LAMOST is operated and managed by the National Astronomical Observatories, Chinese Academy of Sciences. 

This work has made use of data from the European Space Agency (ESA) mission {\it Gaia} (\url{https://www.cosmos.esa.int/gaia}), processed by the {\it Gaia} Data Processing and Analysis Consortium (DPAC, \url{https://www.cosmos.esa.int/web/gaia/dpac/consortium}). Funding for the DPAC has been provided by national institutions, in particular the institutions participating in the {\it Gaia} Multilateral Agreement. 

Funding for the Sloan Digital Sky Survey IV has been provided by the Alfred P. Sloan Foundation, the U.S. Department of Energy Office of Science, and the Participating Institutions. SDSS-IV acknowledges support and resources from the Center for High Performance Computing  at the University of Utah. The SDSS website is www.sdss.org. SDSS-IV is managed by the Astrophysical Research Consortium for the Participating Institutions of the SDSS Collaboration including the Brazilian Participation Group, the Carnegie Institution for Science, Carnegie Mellon University, Center for Astrophysics | Harvard \& Smithsonian, the Chilean Participation Group, the French Participation Group, Instituto de Astrof\'isica de Canarias, The Johns Hopkins University, Kavli Institute for the Physics and Mathematics of the Universe (IPMU) / University of Tokyo, the Korean Participation Group, Lawrence Berkeley National Laboratory, Leibniz Institut f\"ur Astrophysik Potsdam (AIP),  Max-Planck-Institut f\"ur Astronomie (MPIA Heidelberg), Max-Planck-Institut f\"ur Astrophysik (MPA Garching), Max-Planck-Institut f\"ur Extraterrestrische Physik (MPE), National Astronomical Observatories of China, New Mexico State University, New York University, University of Notre Dame, Observat\'ario Nacional / MCTI, The Ohio State University, Pennsylvania State University, Shanghai Astronomical Observatory, United Kingdom Participation Group, Universidad Nacional Aut\'onoma de M\'exico, University of Arizona, University of Colorado Boulder, University of Oxford, University of Portsmouth, University of Utah, University of Virginia, University of Washington, University of Wisconsin, Vanderbilt University, and Yale University. 

This research has made use of NASA's Astrophysics Data System Bibliographic Services.
\end{acknowledgments}

\facilities{Gaia; Kepler; Keck:I (HIRES); LAMOST; Sloan (APOGEE)}

\software{\texttt{astropy} \citep{astropy13, astropy18},
          \texttt{astroquery} \citep{astroquery},
          \texttt{corner} \citep{corner},
          \texttt{dustmaps} \citep{dustmaps}, 
          \texttt{emcee} \citep{emcee2013, emcee2019},
          \texttt{jupyter} \citep{jupyter},
          \texttt{matplotlib} \citep{matplotlib},
          \texttt{numpy} \citep{numpy},
          \texttt{pandas} \citep{pandas-soft, pandas-proc},
          \texttt{scipy} \citep{scipy},
          \texttt{seaborn} \citep{seaborn},
          \texttt{showyourwork!} \citep{Luger2021c}} 

\bibliography{bib}{}
\bibliographystyle{aasjournal}

\end{document}